\documentclass{elsart}

\usepackage{graphicx}
\usepackage{natbib}

\usepackage{graphicx}

\usepackage{amssymb}
\def\be{\begin{equation}}
\def\ee{\end{equation}}

\begin{document}

\begin{frontmatter}


\title{Laws of crack motion and phase-field models of fracture}

\author[label1]{Vincent Hakim}
\address[label1]{Laboratoire de Physique Statistique, 
CNRS-UMR8550 associ\'e aux universit\'es Paris VI et VII,
Ecole Normale Sup\'erieure, 24 rue Lhomond, 75231 Paris, France}
\author[label2]{Alain Karma}
\address[label2]{
Physics Department and Center
for Interdisciplinary Research on
Complex Systems, Northeastern University,
Boston, Massachusetts 02115}


\begin{abstract}
Recently proposed phase-field models offer self-consistent descriptions
of brittle 
fracture. Here, we analyze these theories in the quasistatic regime of
crack propagation. We  show how to derive the laws of crack motion  either 
by using
solvability conditions in a perturbative treatment for slight departure from the 
Griffith threshold, or by generalizing the    
Eshelby tensor to phase-field models.  
The analysis provides a simple physical interpretation of the second 
component of
the classic Eshelby integral in the limit of vanishing crack
propagation velocity: 
it gives the elastic torque on the crack tip that is 
needed to balance the Herring torque arising from the anisotropic 
interface energy. This force balance condition reduces in this limit
to the principle of local symmetry in isotropic media and to the principle of 
maximum energy release rate for smooth curvilinear cracks in anisotropic media.
It can also be interpreted physically in this limit based on
energetic considerations in the traditional framework of 
continuum fracture mechanics, in support of its general validity
for real systems beyond the scope of phase-field models.  
Analytical predictions of crack paths in anisotropic media are validated by numerical simulations.
Simulations also show that these predictions hold even if the phase-field dynamics is
modified to make the failure process irreversible.
In addition, the role of dissipative forces on the process zone scale as well as 
the extension of the results to motion of planar cracks under pure
antiplane shear are discussed.

\end{abstract}

\date{\today}
\end{frontmatter}

\section{Introduction.}
\label{intro.sec}
The prediction of the path chosen by a crack
as it propagates into a brittle material is a fundamental
problem
of fracture
mechanics. It has classically been addressed 
in a theoretical framework where the equations
of linear elasticity are solved with zero traction
boundary conditions on crack surfaces
that extend to a sharp tip \citep{refgen}. In this description,
the stress distributions
near the crack tip  have 
the universal divergent forms 
\citep{Wil1957,Irw1957} 
\be
\sigma_{ij}^m(r,\Theta)=\frac{K_m}{\sqrt{2\pi r}}f_{ij}^m(\Theta),
\label{km}
\ee
where $K_m$ are the stress intensity factors (SIF)
for the three standard modes I, II, or III of fracture
($m=1,2$ or $3$), $\Theta$
is the angle between the radial vector of
magnitude $r$ with
origin at the crack tip and the local
crack direction and the explicit expressions of the $f_{ij}$'s are
recalled in Appendix \ref{sol.app} (see Eq.~(\ref{str}). The allied energy 
release rate 
(or crack
extension force) reads, for plane strain,
\be
G=\alpha(K_1^2+K_2^2)+K_3^2/(2\mu),\label{Gdef} 
\ee
where $\nu$ denotes Poisson's ratio,
$\mu$ is the shear modulus,
and $\alpha\equiv (1-\nu)/(2\mu)$.
Following \citet{Gri1920}, \citet{Irw1957} postulated that
for the crack to propagate, $G$ 
must exceed some material dependent threshold $G_c$
that is theoretically equal to twice the surface energy ($G_c=2\gamma$),
but often larger in practice. 
Like other problems in fracture,
the prediction of  the crack direction of propagation
was first examined \citep{BarChe1961} for mode III
which is simpler because
the antiplane
component of the displacement vector
$u_3$ is a purely scalar Laplacian field. In this case,
the stress distribution
near the tip, can be expanded as
\begin{equation}
\sigma_{3\Theta}\equiv  \frac{\mu}{r}
\frac{\partial u_3}{\partial \Theta}
=\frac{K_3}{\sqrt{2\pi r}}\cos
\frac{\Theta}{2}-\mu A_2 \sin \Theta
+\dots \label{sigma3}
\end{equation}
The dominant divergent contribution is always symmetrical
about the crack direction. As a consequence, the
knowledge of $K_3$ alone
cannot predict any other path than a straight one.
To avoid this impasse,
\citet{BarChe1961} retained the
subdominant $\sin\Theta$ term, which breaks this symmetry. They
hypothesized that a curvilinear crack
propagates along a direction
where $A_2=0$, when the stress distribution
is \emph{symmetrical} about the crack direction.
In subsequent extensions of this work,
several criteria have been proposed
for plane loading, for which
the tensorial nature
of the stress fields
makes it possible to predict non-trivial crack
paths purely from the knowledge of the stress-intensity
factors
\citep{GolSal1974,CotRic1980}.
The generally-accepted condition ``$K_2=0$''
assumes that the crack propagates
in a pure opening mode with a symmetrical stress
distribution about its local axis \citep{GolSal1974} and
is the direct analog
for plane strain ($u_3=0$) of the condition
$A_2=0$ for mode III.
This ``principle of local symmetry'' has been
rationalized using plausible arguments \citep{CotRic1980} but
cannot be fully derived
without an explicit description of the
process zone, where elastic strain energy is both dissipated
and transformed nonlinearly into
new fracture surfaces. As a result, how to extend
this principle to anisotropic materials, where
symmetry considerations have no obvious generalization,
is not clear \citep{Mar2004}.  This is also the case for curved
three-dimensional
fractures although this appears little-noted in the literature.
In addition, path
prediction remains largely
unexplored for mode III even
for isotropic materials. 

Continuum models of
brittle fracture that describe both short scale failure
and macroscopic linear elasticity within a
self-consistent set of equations have recently been proposed
\citep{Aro2000,KKL2001,Eas2002,Wan2002,Mar2005}.
These models
have already 
shown their usefulness in various numerical simulations.
For both antiplane \citep{KarLob2004} and plane
\citep{HenLev2004} loading, they have proven capable
to reproduce the onset of crack propagation at
Griffith threshold as well as dynamical
branching instabilities \citep{KarLob2004}
and oscillatory \citep{HenLev2004} instabilities.
In a quasistatic setting,
this continuous media approach differs in spirit but 
has nonetheless much in common 
with a  variational approach to brittle fracture \citep{Fra98}
proposed to overcome limitations of Griffith theory. This is 
especially apparent when the latter  is implemented numerically \citep{Bou00},
using ideas \citep{Amb90} initially developed for image segmentation
\citep{Mum89}. 

In this article, we analyze these self-consistent theories of brittle fracture
and show how to {\em derive}  laws of motion for the crack tip.
This provides, in particular, relations which generalize the 
principle of local symmetry for an anisotropic material. Furthermore, we
validate these relations by phase-field simulations. This validation is carried
out both for the traditional variational formulation of the phase-field model
with a so-called ``gradient dynamics'', which guarantees that
the total energy of the system, i.e. the sum of the elastic and cohesive energies,
decreases monotonously in time, and for a simple modification of this
dynamics that makes the failure process irreversible. We find that both
formulations yield essentially identical crack paths that are well predicted
by the laws of crack motion derived from the phase-field model. 

For clarity of exposition, the relations derived from the phase-field model
are summarized first in section \ref{over.sec} and interpreted physically in
the context of previous results from the fracture community. This section stresses why the 
second component of the Eshelby configurational force 
perpendicular to the crack axis is both physically meaningful and 
important for the determination of crack paths, even 
though this force has been largely ignored in the fracture mechanics literature since it
was introduced.

Our approach is 
applicable to
a large class of diffuse interface descriptions
of brittle fracture. However,
for clarity of exposition,
we base our derivation on the
phase-field model introduced by \citet{KKL2001}. As recalled in
section \ref{kkl.sec}, in this description,
 the displacement
field is coupled to a single
scalar order parameter or ``phase field'' $\phi$,
which describes a smooth transition in space between unbroken
($\phi=1$) and broken
states ($\phi=0$) of the material. 
We focus on
quasi-static fracture in a macroscopically isotropic elastic
medium with negligible inertial
effects. Material anisotropy is simply included
by making the surface energy $\gamma(\theta)$, dependent on
the orientation $\theta$ of the crack direction with
respect to some underlying crystal axis.
In section \ref{sol.sec}, we analyze the quasi-static motion of a crack,
perturbatively
for small departure from Griffith threshold ($|G-G_c|\ll 1$) and small 
anisotropy. The crack laws of motion are shown to be determined
in a usual manner by solvability conditions, coming from translation invariance 
parallel
and perpendicular to the crack tip axis. 

A different derivation is provided
in section \ref{esh.sec} by generalizing Eshelby tensor \citep{Esh1975}
to phase-field theories. The particular case of motion under pure
antiplane shear is then discussed in section \ref{mo3.sec}. 
Our analytical
predictions are compared with numerical phase-field simulations 
in section \ref{num.sec} where we also examine the sensitivity of the results to
the irreversibility of the failure process. Our conclusions and some further
perspectives of this work are then presented in section \ref{con.sec}. 
Further information on the phase-field model of  \citet{KKL2001}
is provided in Appendix \ref{kkl.app} in the simple context of a stretched 
one-dimensional band.
Details of some of our
calculations are provided in the following appendices \ref{var.app}
and \ref{sol.app}.  A short version of this work has been published in \citep{hk05}.

\section{An overview of the physical picture and main results in the classical
fracture formalism}
\label{over.sec}

\begin{figure}[ht]
\center
\includegraphics[width=8cm]{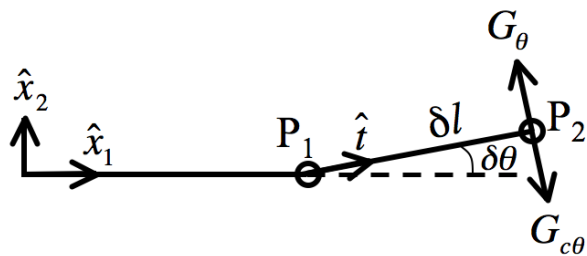}
\caption{Schematic representation of an infinitesimal extension
$P_1P_2$ of the crack of length $\delta l$ at and angle $\delta \theta$ measured with respect to the crack axis. The arrows pointing perpendicular to the crack  denote the two analogs $G_\theta$
and $G_{c\theta}$ of the Herring torque associated with the directional dependence of the crack extension force and the fracture energy around
the crack axis, respectively. 
}
\label{torque}
\end{figure}

In the formalism of continuum fracture mechanics, crack propagation has been
traditionally analyzed by considering the crack extension force $G$ defined by Eq.~(\ref{Gdef}). 
This is a purely configurational force that points along the crack axis in the direction of 
propagation where $G\delta l$ is the amount elastic energy released when the 
crack advances infinitesimally along this axis by a distance $\delta l$. When considering the
propagation of a general curvilinear crack, 
however, it is necessary to consider the extension of a crack at some small 
infinitesimally angle $\delta \theta$ with respect to its current axis as 
depicted schematically in Fig. \ref{torque}.
Physically, one would expect a 
configurational force, distinct from $G$, 
to be associated with the extra amount of elastic energy that is released if the crack propagates by $\delta l$ along this new direction, denoted here by $\hat t$, as opposed to propagating the same distance along its current axis, denoted by $\hat x_1$. 

This additional force on the crack tip was considered by \citet{Esh1975}. It 
can be interpreted physically as producing a torque on the crack tip that changes the crack propagation direction so as to maximize the elastic energy released. The force that produces this torque must act 
perpendicularly to the crack propagation direction and its magnitude is simply 
\begin{equation}
 G_\theta \equiv \lim_{\delta\theta\rightarrow 0}\frac{\delta G}{\delta \theta} \label{Gtorquedef}
\end{equation}
where $\delta G$ is the difference between the crack extension force along the new direction and the old direction, i.e. along $\hat t$ and $\hat x_1$ in Fig. \ref{torque}.  This torque is analogous to the well-known ``Herring torque'' \citep[p. 143]{Her1951} acting on the junction of three crystal grains of different orientations in a polycrystalline material, with the main difference that $G_\theta$ is a configurational force in the present fracture context while the Herring torque is produced by a physical force associated with the grain boundary energy, $\gamma_{gb}(\theta)$,  which is generally anisotropic.  
This force acts perpendicularly to each grain boundary segment at the junction of three grains with a magnitude  $d\gamma_{gb}/d\theta$.

This analogy suggests that there should generally be two torques acting on the crack tip. 
The first, already mentioned, is Eshelby's configurational elastic torque $G_\theta$ associated with the directional dependence of the crack extension force in reference to the local crack axis. The second is the physical torque associated with the directional dependence of the fracture energy, defined here by $G_c(\theta)$, which 
should have a magnitude $dG_c(\theta)/d\theta \equiv G_{c\theta}$ by direct translation of Herring's result for fracture. It follows that the balance of forces at the crack tip should yield two conditions. The first is the standard condition of the classical fracture formalism associated with the balance of forces along the crack axis,
$G=G_c$. 
The second is a new condition
\begin{equation}
G_\theta=G_{c\theta}=2\gamma_{\theta}, \label{torquebalance}
\end{equation}
which corresponds physically to the balance of the two aforementioned torques acting on the crack tip. While $G_\theta$ pulls the crack in a direction that
tends to maximize the release of elastic energy, $G_{c\theta}$ pulls the crack in a direction that minimizes the energy cost of creating new fracture surfaces. 
The second equality on the right-hand-side of Eq. (\ref{torquebalance}) 
only holds in some ideal brittle limit where the fracture energy is  
equal to twice the surface energy, defined here by $\gamma(\theta)$, and $\gamma_\theta\equiv d\gamma(\theta)/\d\theta$. We note that this ideal brittle limit is exact for the class of phase-field models analyzed in this paper but at best only approximate even for a strongly brittle material such as glass. The issue of the quantitative evaluation of $G_{c\theta}$, however, should be kept separate from its role in crack path prediction that is our main focus in this paper.

To see how this torque balance condition provides an explicit prediction for the crack path, it is useful to derive an expression for $G_\theta$ by elementary means, directly from the definition of Eq. (\ref{Gtorquedef}), instead of by evaluating an Eshelby-Rice type integral around the crack tip \citep{Ric1968,Esh1975}, as done later in this paper (see section \ref{esh.sec} and Appendix C); 
while both methods yield the same answer, the former is more physically transparent. For this purpose, we use 
the known expressions for the new stress intensity factors $K_1^*$
and $K_2^*$ at the tip (corresponding to $P_2$ in Fig. \ref{torque}) of
an infinitesimally small kink extension of length $\delta l$ of a semi-infinite crack 
\citep{AmeLeb1992}. In the limit of vanishing kink angle, these
expressions are given by
\begin{eqnarray}
K_1^*&=&  K_1-3K_2\delta \theta/2+\dots 
\label{amel1}
\\
K_2^*&= & K_2+K_1\delta \theta/2+\dots
\label{amel2}
\end{eqnarray}
to linear order in $\delta \theta$ independently of $\delta l$, 
where $K_1$ and $K_2$ are the
stress intensity factors at the tip (corresponding to $P_1$ in Fig. \ref{torque})
of the original straight crack. Using Eq.~(\ref{Gdef}) with
these new stress intensity factors to define
$G(\delta \theta)$, we obtain at once that
$\delta G=G(\delta \theta)-G(0)=-2\alpha K_1K_2\delta\theta$, and hence
using Eq. (\ref{Gtorquedef}), that $G_\theta=-2\alpha K_1K_2$.
Substituting this expression for $G_\theta$ in the torque balance
condition (\ref{torquebalance}), we obtain the condition
\be
K_2=\frac{G_{c\theta}}{2\alpha K_1}=- \frac{\gamma_{\theta}}{\alpha K_1},
\label{k2simpint}
\ee
where second equality only holds in the ideal brittle limit as before. In the isotropic limit where $G_{c\theta}$ vanishes, this condition reduces to the principle of local symmetry which assumes that the crack propagates in a pure opening mode ($K_2=0$). In contrast, for an anisotropic material, $K_2$ is finite with a magnitude that depends both on $K_1$ and the local crack propagation direction, i.e. $G_{c\theta}$ depends on the direction of the crack with respect to some fixed crystal axis in a crystalline material.  
For simplicity, we have restricted our derivation to a situation where linear elasticity is isotropic (e.g., hexagonal symmetry in two dimensions), but Eq. (\ref{k2simpint}) could straightforwardly be extended to a more general situation where linear elasticity is also anisotropic.

The recognition that the torque balance condition (\ref{torquebalance}) can be used to determine the general path of a crack in a brittle material is the central result of this paper. This condition sheds light on the physical origin of the principle of local symmetry in the isotropic limit and shows how it can be generalized quantitatively to anisotropic materials. Although the configurational force perpendicular to the crack tip was considered explicitly by \citet{Esh1975}, is has been largely ignored until recently. This is perhaps because the displacement of a small segment of crack perpendicular to itself, which one might naively expect to result from such a force, would appear unphysical and unreconcilable with the irreversibility of the fracture process. While such a motion is unphysical, it should be clear from the present considerations that all the torques acting on the crack tip, 
both the elastic configurational torque $G_\theta$ and the physical torque  $G_{c\theta}$ linked to fracture energy
anisotropy, have been obtained solely from the consideration of an infinitesimal, physically admissible, extension of the crack at a small angle from its axis. In equating these two torques at the crack tip, the main assumption made is that the dynamics on the process zone scale is able to sample different possible microscopic states so as to permit local relaxation to mechanical equilibrium. 

There have been more recent attempts to incorporate the Eshelby elastic torque in the classical fracture formalism, where fracture surfaces are treated as mathematically sharp boundaries extending to the crack tip \citep{Add1999,Ole2001,Mar2004}. These theories, however, have not produced an explicit torque balance condition analogous to Eq. (\ref{k2simpint}) that can be formally derived and tested. From this standpoint, the phase-field framework has the advantage of removing many of the ambiguities that arise when considering the motion of the crack tip in the classical fracture formalism.
In the phase-field framework, a torque balance condition can be rigorously derived from the condition for the existence of a propagating crack solution that is spatially diffuse on the inner scale of the process zone, and must match smoothly to the standard solution of linear elasticity on the outer scale of the sample size. This condition reduces to Eq.  (\ref{torquebalance}) (or Eq. (\ref{k2simpint}) for isotropic elasticity) in the limit of vanishing crack velocity, but contains additional contributions for finite crack velocity associated with dissipative forces on the process zone scale. 

Interestingly, the results of the phase-field analysis show that the component of the dissipative force perpendicular
to the crack tip vanishes for propagation in isotropic media because both the stress distribution and the 
phase field are symmetrical about the crack axis in this case.
Consequently, within the phase-field framework, dissipative forces do 
not change the condition $K_2=0$ for crack propagation in isotropic media. For propagation in anisotropic media,
in contrast, small velocity-dependent correction to the 
torque balance condition (\ref{k2simpint}) arise because this symmetry is broken.

\section{The KKL phase-field model of fracture}
\label{kkl.sec}
Fracture is generally described in diffuse interface models
\citep{Aro2000,KKL2001,Eas2002,Wan2002,Mar2005} as a softening
of the elastic moduli at large strains. This can be done purely
in term of the strain tensor but it produces field equations with
derivative of high order \citep{Mar2005}. Here, we adopt the alternative
approach of introducing a supplementary field $\phi$, a scalar order
parameter or ``phase-field'', that describes
the state of the material and smoothly interpolates between intact
($\phi=1$) and fully broken ($\phi=0$) states. For definiteness,
we base our derivation on the specific model proposed by \citet{KKL2001}
with energy density $\mathcal{E}$,
\begin{equation}
\mathcal{E}= \mathcal{E}_{pf}(\{\partial_j \phi\})+
g(\phi) (\mathcal{E}_{strain}-
\mathcal{E}_c) + \mathcal{E}_c
\end{equation}
where $\partial_j\equiv \partial/\partial x_j$ denotes the partial derivative with
respect to the cartesian coordinate 
$x_j$ ($j=1,2,3$) and $\mathcal{E}_{strain}$ is the elastic energy of the intact material.
The equations of motion are derived variationally from the total energy  
of the system that is the spatial integral 
\begin{equation}
 E=\int d^3x ~\mathcal{E} 
 \end{equation}
of the energy density.
In the quasitatic case, these are
\begin{eqnarray}
0&=&
-\frac{\delta E}{\delta u^k}=\partial_j
\frac{\partial {\mathcal E}}{\partial
[\partial_j u^k]}-\frac{\partial {\mathcal E}}{\partial u^k}
\label{varu.eq}\\
\chi^{-1}\partial_t\phi&=&
-\frac{\delta E}{\delta \phi}=\partial_j
\frac{\partial {\mathcal E}}{\partial
[\partial_j \phi]}-\frac{\partial {\mathcal E}}{\partial \phi}
\label{varphi.eq}
\end{eqnarray}

The three Euler-Lagrange Eq.~(\ref{varu.eq}) for the 
cartesian components
$u_k$ of the displacement vector ($k=1,2,3$) are
simply the static equilibrium conditions
that the sum
of all forces
on any material
element
vanish.
The fourth Eq.~(\ref{varphi.eq}) for $\phi$
is the standard Ginzburg-Landau form that governs
the phase-field evolution, with
$\chi$  a kinetic coefficient that controls the rate
of energy dissipation in the process zone, i.e. it follows from Eqs. (\ref{varu.eq})  and
(\ref{varphi.eq}) that
\begin{equation}
\frac{dE}{dt}=-\chi \int d^3x \left(\frac{\delta E}{\delta \phi} \right)^2\le 0\label{gradyn}
\end{equation}

In the simplest case of an isotropic elastic medium and isotropic $\phi$,
the phase-field and strain energy are simply,
\begin{eqnarray}
\mathcal{E}_{pf}(\{\partial_j \phi\})&=& \frac{\kappa}{2} (\nabla \phi)^2
\label{eniso.def}
\\
\mathcal{E}_{strain}(\{u_{ij}\})&=& \frac{\lambda}{2} (u_{ii})^2 +\mu u_{ij} u_{ij}
\label{eliso.def}
\end{eqnarray}
where $u_{ij}=(\partial_i u_j+\partial_j u_i)/2$ is the usual strain
tensor of linear elasticity.
No asymmetry between
dilation and compression is included since this is
not necessary for our present purposes.
The broken state of the material becomes
energetically favored when ${\mathcal E}_{\rm strain}$
exceeds the threshold ${\mathcal E}_c$
and $g(\phi)$ is a monotonically increasing
function of $\phi$ that describes the softening of the elastic
energy at large strain ($g(0)=0$) and produces the usual elastic behavior
for the intact material ($g(1)=1, g'(1)=0$). In addition, the
release  of bulk stress by a crack
requires the function $g(\phi)$  to vanish
faster than $\phi^2$ for small $\phi$, as recalled in
Appendix \ref{kkl.app}.
We therefore choose $g(\phi)=4\phi^3 - 3\phi^4$, as in 
\citep{KKL2001,KarLob2004,HenLev2004}.
With these choices, the isotropic interface energy is equal to
\be
\gamma_0= \sqrt{2\kappa\mathcal{E}_c}\int_0^1 d\phi\sqrt{1-
 g(\phi)}\simeq 0.7165 \sqrt{2\kappa\mathcal{E}_c}
\label{stiso}
\ee
as shown
in Appendix \ref{kkl.app} (Eq.~(\ref{eexpmp2})),
by repeating the analysis of \citet{KKL2001}.

In the present paper, we analyze the case of a phase-field energy
$\mathcal{E}_{pf}(\{\partial_j \phi\})$ without rotational symmetry
which gives an anisotropic interface energy. A simple example used
for concreteness and for the numerical simulations is provided by
a simple two-fold anisotropy in the phase-field energy
\footnote{Note that with  coordinates $x',y'$ rotated by $\pi/4$ with respect
to the $x,y$ axes, the phase-field energy reads,
$\
\mathcal{E}_{pf}=  \frac{\kappa}{2}(1+\epsilon/4)(\partial_{x'}\phi)^2
+\frac{\kappa}{2}(1-\epsilon/4) (\partial_{y'}\phi)^2,\
$.
}
\be
{\mathcal E}_{pf}=\frac{\kappa}{2}\left(|\nabla \phi|^2
+ \epsilon \partial_1\phi\partial_2\phi \right)
\label{enani.def}
\ee
The interface energy
of a straight fracture interface oriented at an angle $\theta$ with
the $x$-axis arises from the variation of the elastic and phase fields
in a direction transverse to the fracture, namely with $\phi(x,y)=\phi[-x 
\sin(\theta)
+y \cos(\theta)]$. Therefore, 
the only difference between Eq.~(\ref{eniso.def}) and
Eq.~(\ref{enani.def}) in this one-dimensional calculation
of the interface energy (Appendix \ref{kkl.app}) 
is the replacement of $\kappa$ by $\kappa[1-(\epsilon/2)\sin 2\theta]$
in the anisotropic case.
The allied anisotropic interface energy thus follows directly from
the isotropic expression (\ref{stiso}) and reads
\be
\gamma(\theta)=\gamma_0\sqrt{1-(\epsilon/2)\sin 2\theta}
\label{stani2}
\ee
It reduces of course to the isotropic surface energy $\gamma_0$
of Eq.~(\ref{stiso}) in the
$\epsilon\rightarrow 0$ limit.

With the specific energies of Eq.~(\ref{eliso.def}) and 
(\ref{enani.def}), the variational phase-field equations
read
\begin{eqnarray}
\partial_j[\sigma_{ij} g(\phi)]&=& 0\nonumber\\
\kappa [\nabla^2 \phi +\epsilon \partial_{xy}\phi]
-g'(\phi)(\mathcal{E}_{strain}-\mathcal{E}_c)&=&\frac{1}{\chi}\partial_t\phi
\label{pfe}
\end{eqnarray}
where $\sigma_{ij}$ is the usual stress tensor for an isotropic medium
\be
\sigma_{ij}=\lambda u_{kk} \delta_{i,j} + 2 \mu u_{ij}
\label{st.def}
\ee
Our aim in the following sections is to analyze the laws that govern
the motion of a crack tip in this
 self-consistent description.

\section{Laws of crack motion as solvability conditions}
\label{sol.sec}
\subsection{The tip inner problem}
In the phase-field description, the obtention
of laws of motion for a crack tip can be viewed
as an ``inner-outer'' matching problem. The phase-field Eq.~(\ref{pfe}) 
introduce an intrinsic process zone scale $\xi=\sqrt{\kappa/(2\mathcal{E}_c)}$.
The ``inner'' problem consists in the
determination of
a solution of Eq.~(\ref{pfe}) at the process zone scale $\xi$.
The
boundary conditions on this inner problem are
imposed at a distance from the crack tip much greater
than the process zone scale ($r\gg\xi$) and much smaller than any
macroscopic length.  They should coincide with
the short-distance asymptotics of 
the ``outer'' problem, namely the usual determination of the 
elastic field for the
crack under consideration. Therefore,
the imposed boundary conditions on Eq.~(\ref{pfe}) are
that the material is intact ($\phi\rightarrow 1$) away from the crack 
itself, and that
for  mixed mode I/II conditions, the asymptotic behavior 
of the displacement field
is
\begin{equation}
u_i(r,\Theta)\sim \frac{1}{4\mu}\sqrt{\frac{r}{2\pi}}[K_1\, d_i^I(\Theta;\nu)+
K_2\, d_i^{II}(\Theta;\nu)]
\label{bcs}
\end{equation}
where $\mu$ is the shear modulus and
the functions $d_i^m$ are directly related to the universal divergent forms
of the stress (Eq.~\ref{km}) and are explicitly given 
(in polar coordinates) by Eq.~(\ref{u1}) and
(\ref{u2}) of Appendix \ref{sol.app}. The values of $K_1$ and $K_2$ are
imposed by the boundary conditions at the macroscopic scale and do not
significantly vary when the crack tip advances by a distance of order $\xi$. In the
frame of the crack tip moving at velocity $v$, 
Eqs.~(\ref{pfe}) thus read
\begin{eqnarray}
\partial_j[\sigma_{ij} g(\phi)]&=& 0
\label{pfes}
\\
\kappa \nabla^2 \phi
-g'(\phi)(\mathcal{E}_{strain}-\mathcal{E}_c)&=&-\frac{v}{\chi}\partial_x\phi
-\epsilon\kappa\,\partial_{xy}\phi
\nonumber
\end{eqnarray}

\subsection{Perturbative formalism and solvability conditions}

Our first approach for obtaining the laws of crack tip motion consists in 
analyzing the  slowly moving solutions of Eq.~(\ref{pfes}) with boundary
conditions (\ref{bcs}) perturbatively 
around an immobile Griffith crack. For isotropic elastic 
and phase field energies and a pure opening mode, this Griffith crack
corresponds to the stationary
solution that exists for $\alpha (K_1^c)^2=G_c$.
Accordingly, we
consider, for a small departure from Griffith threshold,
$\delta K_1=|K_1-K_1^c|/K_1^c\ll 1$ and for a small $K_2\ll K_1^c$,
a slowly moving crack with a small (two-fold) anisotropy in
$\phi$-energy (Eq.~(\ref{stani2})).
Our
 aim is to find the relations between $K_2$ and
the anisotropy, as well as between $K_1, K_2$ and the velocity $v$, required
for the solution existence. 

Linearization of Eqs. (\ref{pfes}) around the isotropic Griffith crack
$u_i^{(0)},\phi^{(0)}$
with the substitutions $u_i=u_i^{(0)}+u_i^{(1)},\phi=\phi^{(0)}+\phi^{(1)}$,
gives,
\begin{eqnarray}
\partial_j[\sigma^{(1)}_{ij} g(\phi^{(0)})]+
\partial_j[\sigma^{(0)}_{ij} g'(\phi^{(0)})\phi^{(1)}]=0
\label{linl}
\\
\kappa \nabla^2 \phi^{(1)}
-g'(\phi^{(0)})\sigma_{ij}^{(0)}u_{ij}^{(1)}
-g''(\phi^{(0)})\phi^{(1)}[\mathcal{E}_{strain}-\mathcal{E}_c]&=&
-\frac{v}{\chi}\partial_x\phi 
-\epsilon\kappa \partial_{xy}\phi
\nonumber
\end{eqnarray}
This can symbolically be written as
\begin{equation}
\mathcal{L}\left(\begin{array}{c}u_1^{(1)}\\u_2^{(1)}\\ \phi^{(1)}\end{array}
\right)=-\frac{v}{\chi}\left(\begin{array}{l}0\\0\\ \partial_x\phi^{(0)}
\end{array}\right)
-\epsilon\kappa\left(\begin{array}{l}0\\0\\
\partial_{xy}\phi^{(0)}\end{array}
\right)
\label{lineq}
\end{equation}
where $\mathcal{L}$ is the linear operator on the left-hand-side (l.~h.~s.~)
of Eq.~(\ref{linl}).
The boundary conditions at infinity are that $\phi^{(1)}$ vanishes and that
$u^{(1)}$ behaves asymptotically as in Eq.~(\ref{bcs}) but 
with $K_1$ replaced by $\delta K_1$, the
small departure from  Griffith threshold, and $K_2$ is also assumed to
be small.

The linear operator  $\mathcal{L}$ 
possesses two right zero-modes, 
that arise
from the invariance of the zeroth-order problem
under $x$ and $y$ translations, and can be explicitly
obtained by infinitesimal translation
of the immobile Griffith crack.
For a general linear operator, the determination
of the left zero-modes would nonetheless be a difficult problem.
However, the variational
character of the equations of motion imposes quite generally that
$\mathcal{L}$ is self-adjoint (see Appendix \ref{var.app}) and that
left zero-modes are identical to right zero-modes. Thus, taking
the scalar product of the two sides of Eq.~(\ref{lineq})
with the two translation zero modes provides two explicit 
solvability conditions 
for Eq.~(\ref{lineq}).

The scalar product with a  left zero-mode $(u_1^L,u_2^L,\phi^L)$ can 
generally be written
\begin{equation}
\int\!\!\!\int dx dy \left(u_1^L, u_2^L, \phi^L\right)\mathcal{L}\left(\begin{array}{c}u_1^{(1)}\\u_2^{(1)}\\
 \phi^{(1)}\end{array}
\right)= -\int\!\!\!\int dx dy\ \phi^L\,\{\frac{v}{\chi} \partial_x\phi^{(0)}
+ \epsilon\kappa
\partial_{xy}\phi^{(0)}\}
\label{scg}
\end{equation}
Since the left vector is a zero mode of $\mathcal{L}$,
the only contribution to the l.~h.~s.~ of Eq.~(\ref{scg}) 
comes from
boundary terms,
\begin{eqnarray}
\int\!\!\!\int dx dy \left(u_1^L, u_2^L, \phi^L\right)
\mathcal{L}
\left(\begin{array}{c}u_1^{(1)}\\u_2^{(1)}\\
 \phi^{(1)}\end{array}
\right)&=&
\label{scgr}\\
\oint  ds~ n_j\left \{
[u_i^L\sigma_{ij}^{(1)}-
u_i^{(1)}\sigma_{ij}^L]\ g(\phi^{(0)})
\right.&+&
\left.
[u_i^L\phi^{(1)}-u_i^{(1)}\phi^L]\
g'(\phi^{(0)})\sigma_{ij}^{(0)}
+\kappa [\phi^L\partial_i\phi^{(1)}-\phi^{(1)}\partial_i\phi^L]\right\}
\nonumber
\end{eqnarray}
where $\mathbf{n}$ is the outward contour normal and the contour integral is taken counterclockwise along a  circle (of radius $r$) centered on
the fracture tip.

\subsection{
Translations along $x$ and crack velocity}
The zero mode corresponding to translations along x is
$(\partial_x u_1^{(0)},\partial_x u_2^{(0)},\partial_x \phi^{(0)}
)$.
On the right-hand-side (r.~h.~s.) of Eq.~(\ref{scg}), the term 
proportional
to the anisotropy $\epsilon$ 
vanishes (by symmetry or explicit integration).
The  r.~h.~s. of Eq.~(\ref{scgr}) can be simplified since
on a circle of a large
enough radius, $g(\phi^{(0)})$ equals unity everywhere
except in the region where the circle
cuts the fracture lips. This region of non constant $\phi$ is far away
from the crack tip where the
crack is to a very good approximation invariant by translation along x and
$\partial_x \phi\simeq\partial_x \mathbf{u}\simeq 0$. Therefore,

\begin{eqnarray}
\int\!\!\!\int dx dy \left(\partial_x u_1^{(0)},\partial_x u_2^{(0)},\partial_x \phi^{(0)}
\right)\mathcal{L}\left(\begin{array}{c}u_1^{(1)}\\u_2^{(1)}\\
 \phi^{(1)}\end{array}
\right)&=&
\oint ds \ n_i [u_i^{(x;I)}\sigma_{ij}^{(1)}-
u_i^{(1)}\sigma_{ij}^{(x;I)}]
\nonumber\\
&=&
-\frac{K_1\,\delta K_1}{\mu}(1-\nu)
\label{scxsimp}
\end{eqnarray}
where the explicit 
formulas (\ref{u1},\ref{u2})
for the elastic displacements around a straight crack,
have been used to 
obtain the last equality
as detailed in Appendix
\ref{sol.app} (see Eq.~(\ref{normx})).
Comparison between Eq.~(\ref{scxsimp}) and Eq.~(\ref{scg})
finally provides the natural result
that the crack velocity 
is proportional to
the departure from  Griffith threshold,  
\begin{equation}
\frac{v}{\chi}\int\!\!\!\int dx dy [\partial_x\phi^{(0)}]^2=
\frac{K_1\,\delta K_1}{\mu}(1-\nu)
=\delta G
\end{equation}

\subsection{
Translations along $y$ and crack direction}
A second condition on crack motion arises from
the zero mode corresponding to translations along y,
$(\partial_y u_1^{(0)},\partial_y u_2^{(0)},\partial_y \phi^{(0)})$.
In this case, only the term proportional
to 
the anisotropy $\epsilon$ 
contributes to the l.~h.~s.~ of
Eq.~(\ref{scgr}).
\begin{equation}
\int\!\!\!\int dx dy\, \partial_y\phi^{(0))} \partial_{xy}\phi^{(0))}
= -
\int dy [\partial_y\phi^{(0))}|_{x=-\infty}]^2
\end{equation}

Similarly to Eq.~(\ref{scxsimp}),
the  r.~h.~s. of Eq.~(\ref{scgr}) simplifies  when the integration
contour is a large enough circle 

\begin{eqnarray}
\int\!\!\!\int dx dy \left(\partial_y u_1^{(0)},\partial_y u_2^{(0)},\partial_y \phi^{(0)}
\right)\mathcal{L}\left(\begin{array}{c}u_1^{(1)}\\u_2^{(1)}\\
 \phi^{(1)}\end{array}
\right)&=&
\oint ds \ n_i [u_i^{(y;I)}\sigma_{ij}^{(1)}-
u_i^{(1)}\sigma_{ij}^{(y;I)}]\nonumber\\
&=&\frac{K_1\, K_2}{\mu}(1-\nu)
\label{scysimp}
\end{eqnarray}
where again the explicit evaluation in the last equality is detailed in
Appendix \ref{sol.app} (see Eq.~(\ref{normy}).
Thus, the second relation of crack motion reads
\begin{equation}
\frac{K_1\, K_2}{\mu}(1-\nu)=
\frac{\epsilon\kappa}{2} 
\int_{-\infty}^{+\infty}\!\!\! dy\,
\left[\partial_y\phi^{(0))}|_{x=-\infty}\right]^2
\label{eqcm2p}
\end{equation}
Eq.~(\ref{eqcm2p}) reduces to the principle of local symmetry (i.~e.~ $K_2=0$)
for an isotropic medium and provides the appropriate generalization for the
considered anisotropy.
Before further discussing these results and their physical consequences,
we present a different derivation in the next section.

\section{Generalized Eshelby-Rice integrals}
\label{esh.sec}

The second approach, which we pursue here, directly
exploits the variational
character of the equations of motion and their invariance under translation.
It yields identical solvability conditions
as the approach of section \ref{sol.sec} when $G-G_c$ and
symmetry breaking perturbations are small, but it is
more general since it does not require
these quantities to be small.

\subsection{The Generalized Eshelby tensor}
As shown by E. Noether in her classic work \citep{Noether}, 
to each continuous symmetry of variational equations is
associated 
a conserved
quantity (charge) and an allied divergenceless current.
Space (and time) translation invariance are well-known
to give the divergenless energy-momentum
tensor in field theories \citep{LLtf}.
\citet{Esh1951} and following authors 
\citep{Ric1968,Esh1975,Gur1998,Add1999,Ole2001}
 have shown the usefulness of 
the analogous tensor for classical elasticity theory.
Here, we consider
the generalized
energy-momentum (GEM) tensor which
extends  Eshelby tensor
for linear elastic fields \citep{Esh1975} by
incorporating
short-scale physics through its additional dependence
on the phase-field $\phi$.

We find it convenient to define the four-dimensional vector field
$\psi^{\alpha}=u_{\alpha}$ for $1\le \alpha \le 3$ and
$\psi^{\alpha}=\phi$ for $\alpha=4$,
where $u_{\alpha}$ are the components
of the standard displacement field.
The inner problem Eq.~(\ref{pfes}) can then be rewritten in the condensed form
\be
- \delta_{\alpha,4}\, v\, \chi^{-1} \partial_1 \phi =
\partial_j
\frac{\partial {\mathcal E}}{\partial [
\partial_j \psi^{\alpha}]}-\frac{\partial {\mathcal E}}{\partial \psi^{\alpha}
},\ \ \ \alpha=1,\cdots,4.
\label{eqmo2}
\ee
where here and in the following
summation is implied on repeated indices (from
$1$ to $3$ on roman indices and from $1$ to $4$ on greek ones).
Chain rule differentiation provides the simple equality,
\be
\partial_i {\mathcal E}
=\frac{\partial {\mathcal E}}{\partial
\psi^{\alpha}}\partial_i\psi^{\alpha}
+\frac{\partial{\mathcal E}}{\partial [ \partial_j\psi^{\alpha}]}
\partial_j\partial_i\psi^{\alpha} 
\label{chain}
\ee
Using Eq.~(\ref{eqmo2}) to eliminate $\partial {\mathcal E}/\partial 
\psi_{\alpha}$
from 
the r.h.s. of Eq.~(\ref{chain}),
we obtain
\be
\partial_j\,T_{ij}=\frac{v}{\chi}
\partial_1\phi\partial_i \phi 
~~{\rm for}~~i=1,2.\label{div}
\ee
where the generalized
energy-momentum (GEM) tensor $T_{ij}$ reads
\be
T_{ij}\equiv {\mathcal E}\delta_{ij}
-\frac{\partial{\mathcal E}}{\partial [\partial_j\psi^{\alpha}]}
\partial_i \psi^{\alpha}
\label{gem.def}
\ee
The GEM tensor $T_{ij}$
is the sought extension of
the classical Eshelby tensor $T_{ij}^E$
\citep{Esh1951} of classical linear elasticity
\be
T^E_{ij}= \mathcal{E}_{strain} \delta_{ij}-\sigma_{jk} \partial_i u_{k}
\label{eshE}
\ee
The GEM tensor $T_{ij}$
 reduces identically to $T^E_{ij}$ in the intact material where 
the phase-field is constant ($\phi=1$). Both tensors are
non-symmetric in their two
indices. The divergence of the GEM tensor
taken on its second indice vanishes in the
zero-velocity limit, when dissipation in the process zone also vanishes.

\subsection{Laws of crack motion}
\begin{figure}[ht]
\begin{center}
\includegraphics[width=5cm]{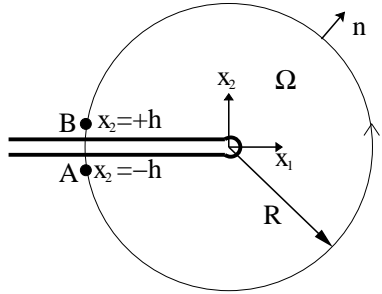}
\end{center}
\caption{Spatially diffuse crack tip region with $\phi=1/2$
contour separating broken and unbroken material (thick solid line).
}
\label{schem}
\end{figure}

In order to take advantage of Eq.~(\ref{div}), we integrate
the divergence
of the GEM tensor over a large disk 
$\Omega$ centered on the crack tip (see Fig.~\ref{schem}), following
Eshelby computation of the
configurational force on the crack tip
treated as a defect in a linear elastic
field \citep{Esh1975} and subsequent
attempts to derive criteria
for crack propagation and stability
\citep{ Gur1998,Add1999,Ole2001}. The important difference with these
previous computations is that, here,
the GEM tensor (\ref{gem.def}) is well-defined everywhere,
so that the crack itself is included in the domain of integration.
%
The integral of the divergence
of the GEM tensor can be written 
as a contour integral over the large circle $\partial \Omega$ bounding
the disk $\Omega$,
\be
F_i=\int_{C_{A\rightarrow B}}\!\!\!\!\! ds \,T_{ij}\,n_j
+\int_{B\rightarrow A}\!\!\!\!\! ds \,T_{ij}\,n_j -\frac{v}{\chi}
\int_\Omega d\vec x~\partial_1\phi\partial_i \phi=0.\label{line}
\ee
We have decomposed the circle $\partial\Omega$
into: (i) a
large loop $C_{A\rightarrow B}$ around the tip in the unbroken
material, where $A$ ($B$) is at a height $h$ below (above)
the crack axis that is much larger than the process zone size but
much smaller than the radius $R$ of the contour, $\xi \ll h \ll R$,
and (ii)
the segment $(B\rightarrow A)$
that traverses the crack from $B$ to $A$ behind the tip,
as illustrated in Fig.~\ref{schem}. In both integrals,
$ds$ is the contour arclength element
and
$n_j$ the components of its outward normal.

Eq.~(\ref{line}) provides an alternative basis to predict the crack
speed and its path for quasi-static fracture. The
$F_i$'s can be interpreted as the 
parallel ($i=1$) and perpendicular ($i=2$) components with respect
to the crack direction,
of the sum of all forces acting on the crack tip.
In Eq.~(\ref{line}),
the three integrals terms from left to right 
respectively
represents
configurational,
cohesive, and dissipative forces. We examine them in turn.

\subsubsection{Configurational forces and Eshelby torque}
We take $A$ and $B$  far back from the tip and close
to the crack on a macroscopic scale but with
the distance $h$ between $A$ and $B$ much larger than the
process zone scale. Namely, we consider the mathematical limit
$h\rightarrow +\infty, R\rightarrow +\infty\ $ with
$h/R\rightarrow 0$ where $R$ is
the distance from $A$ and $B$ to the crack tip. 
In this limit, the first
integral in
Eq.~(\ref{line}) is taken on a path that is entirely in the
unbroken material where $\phi$ is constant and equal to unity. Thus,
the tensor $T_{ij}$ reduces  to the classical
Eshelby 
tensor $T^E_{ij}$ (Eq.~(\ref{eshE}))
the first integral in
Eq.~(\ref{line}) yields the two components of the usual
configurational forces $F_i^{(conf)}$,
\be
F_i^{(conf)}=\int_{C_{A\rightarrow B}} \!\!\! ds \,T^E_{ij}\,n_j
\label{fconf}
\ee

The first component, $F_1^{(conf)}$, is the crack extension
force and also Rice's $J$ integral \citep{Ric1968}.
\be
F_1^{(conf)}=\int_{C_{A\rightarrow B}} \!\!\! ds \,T^E_{1j}\,n_j,\label{f21}\\
\ee
With the known forms of the elastic
displacement fields near the crack tip, as detailed in Appendix \ref{sol.app}
(see Eq.~(\ref{f21eapp})), one obtains
the well-known expression
(\ref{Gdef}) of the crack extension force,
\be
F_1^{(conf)}=G=\alpha(K_1^2+K_2^2),\label{f21e}
\ee

The second component $F_2^{(conf)}$ can be computed in an analogous way 
from the elastic displacement fields near the crack tip,
(Eq.~(\ref{f22eapp}) and one obtains
\be
F_2^{(conf)}
=-2\alpha K_1 K_2,  \label{f22e}
\ee
As discussed earlier, $F_2^{(conf)}$ is the Eshelby torque 
\citep{Esh1975} that is readily interpreted physically
if one imagine
extending the crack tip by a small amount at a small angle $\theta$ from
the main tip axis.
Then
$F_2^{(conf)}$ is equal to the angular derivative of the crack
extension force $G(\theta)$ at $\theta=0$.
\begin{figure}[ht]
\begin{center}
\includegraphics[width=7cm]{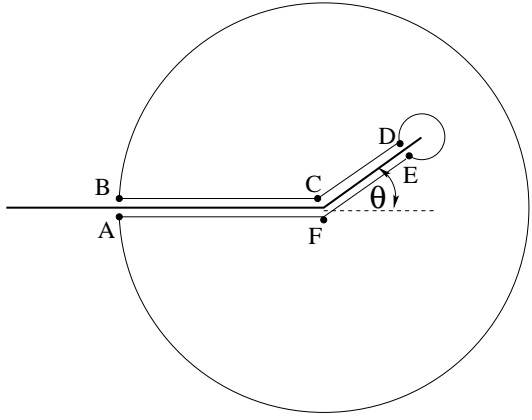}
\end{center}
\caption{Sketch showing the contour integral decomposition in 
Eq.~(\ref{intild}). The crack with its virtual extension at an angle
$\theta$ is depicted by the the thick bold line. The integral contour
follows the great circle from A to B; it continues along the upper
lip
of the crack from B to C and then along the upper lip of the virtual extension
from C to D; it then encircles the extended crack tip following the small 
circle from D to E; finally it comes back to A along the lower crack lips 
via E and F.}
\label{schem2}
\end{figure}

This equality can be seen in two ways. First, we can use the general properties
of Eshelby tensor.
We denote with a tilde the elastic quantities corresponding to
the crack with the small extension of length $s$ at an angle $\theta$.
 Since the crack extension is along the direction $(\cos(\theta),\sin(\theta))$,
we consider the allied
vector obtained from the Eshelby tensor,
 $\tilde{T}^E_{\theta j}\equiv 
\cos(\theta) \tilde{T}^E_{1j}+\sin(\theta)\tilde{T}^E_{2j}$. 
The flux of this vector
vanishes when taken through the contour
that goes along the great circle from $A$ to $B$
and then continues in a classical way along the lips of the extended crack,
as drawn in Fig.~\ref{schem2}
\be
(\int_{C_{A\rightarrow B}} +
\int_{{B\rightarrow C}}+ 
\int_{C\rightarrow D}+
\int_{C_{D\rightarrow E}}+
\int_{E\rightarrow F}+ 
\int_{F\rightarrow A} \!\!\!) ds \,\tilde{T}^E_{\theta j}\,n_j=0
\label{intild}
\ee
The two integrals on the fracture lips  from $C$ to $D$ and $E$ to $F$
do not contribute since the integrand vanishes: the $\theta$ direction is along the path and the
normal stresses vanish on the fracture lips. The same argument shows that the 
integrand is simply equal to $\pm\tilde{\mathcal{E}}_{strain}\sin(\theta)$ for
the integrals from $B$ to $C$ and $F$ to $E$ along the lips of the original fracture.
The integral on the small circle around the extended crack tip
$C_{D\rightarrow E}$ is equal to $-\tilde{G}(\theta)$ where $\tilde{G}(\theta)$ is the
energy release rate at the end of the small crack extension.
Eq.~(\ref{intild}) thus reduces to
\be
\int_{C_{A\rightarrow B}} [\cos(\theta) \tilde{T}^E_{1j}+
\sin(\theta)\tilde{T}^E_{2j}]\,n_j= \tilde{G}(\theta)+\int_R^0 dx \sin(\theta)
[\tilde{\mathcal{E}}^+_{strain}(x)-\tilde{\mathcal{E}}^-_{strain}(x)]
\label{intild1}
\ee
where $\tilde{\mathcal{E}}^+_{strain}$ and $\tilde{\mathcal{E}}^-_{strain}$
respectively denote the elastic strain energy densities on the upper 
and lower fracture lips. The required identity between $F_2^{(conf)}$ and
the angular derivative of $d\tilde{G}/d\theta|_{\theta=0}$ follows from differentiation
of Eq.~(\ref{intild1}) with respect to $\theta$ at $\theta=0$. 
When this is performed, there are two
kinds of terms. Terms coming from the differentiation of the explicit 
trigonometric functions in Eq.~(\ref{intild1}) and terms coming from
the implicit dependence upon $\theta$ of tilde quantities. However, in the integral on the
l.h.s of Eq.~(\ref{intild1}), these implicit terms vanish as the length $s$
of the extension is taken to zero, and in the integral on the r.h.s. they
are multiplied by a vanishing sine function. Moreover, for the straight 
fracture at $\theta=0$, only $\sigma_{xx}$ is non-zero on the fracture lips
and it is of opposite sign on the upper and lower fracture lips 
(Eq.~(\ref{str})). The strain energy densities which are quadratic in the stress $\sigma_{xx}$ are 
therefore equal on the upper and lower fracture lips and after differentiation,
the contribution of integral term on the r.h.s of Eq.~(\ref{intild1}) vanishes
at zero \footnote{Subdominant terms, coming 
for instance from a macroscopic curvature of the crack, could be different on
the two crack lips but note that the
integral range is on a length scale that is
vanishingly small on a macroscopic scale}. Finally, in the limit of a vanishing extension length 
($s\rightarrow 0$) tilde quantity tend toward their (non-tilde) values on the
original fracture and one obtains
\be
F_2^{(conf)}=
\int_{C_{A\rightarrow B}} 
T^E_{2j}\,n_j= \lim_{s\rightarrow 0}\frac{d\tilde{G}(\theta)}{d\theta}|_{\theta=0}\equiv G_{\theta}(0)
\label{intild2}
\ee
This relation between the second component
$F_2^{(conf)}$ and the angular derivative of $G(\theta)$ can also be obtained
by comparing their explicit expressions
in term of the SIF $K_1$ and $K_2$.
As it is well known, the SIF $\tilde{K}_1$ and $\tilde{K}_2$ at the end
of a small extension can be expressed as linear combination of $K_1$ and $K_2$
\citep{AmeLeb1992}
\begin{eqnarray}
\tilde{K}_1&=& F_{11}(\theta) K_1 +F_{12}(\theta) K_2\nonumber\\
\tilde{K}_2&=& F_{21}(\theta) K_1 +F_{22}(\theta) K_2
\label{ktilde}
\end{eqnarray}
with  clearly $F_{11}(0)=F_{22}(0)=1$ and $F_{12}(0)=F_{21}(0)=0$
and the derivative at $\theta=0$,
$F_{11}'(0)= F_{22}'(\theta)=0$, as already mentionned in section \ref{over.sec} (
Eq.~(\ref{amel1},\ref{amel2})).
 A detailed computation \citep{AmeLeb1992}
 provides the
other two derivatives
$F_{12}'(0)=-3/2$ and $F_{21}'(0)=1/2$.
Therefore, one obtains 
\be
\lim_{s\rightarrow 0}\frac{d\tilde{G}(\theta)}{d\theta}_{|\theta
=0}=2\alpha K_1 K_2 [F_{12}'(0)+F_{21}'(0)]=-2\alpha K_1 K_2
\label{dgdcal}
\ee
This is indeed identical to the expression
of $F_2^{(conf)}$ obtained by
a direct computation (Eq (\ref{f22e})) and it provides a
second derivation of
Eq.~(\ref{intild2}).


\subsubsection{Cohesive forces}

An important new ingredient in Eq.~(\ref{line}) is
the second portion of the line integral ($\int_{B\rightarrow A}$)
of the GEM tensor that traverses the crack.
This integral represents physically the
contributions of cohesive forces inside the
process zone. To see this, we first note that
the profiles of the phase-field and the three components of the
displacement can be made to
depend only on $x_2$ provided that the contour is chosen much larger
than the process zone size and to traverse the crack perpendicularly
from $B$ to $A$. With this choice, we have that $n_1=-1$,
$n_2=0$,
along this contour and therefore
that, for $i=1$
\be
F_1^{(coh)}=\int_{B\rightarrow A} ds \,T_{1j}\,n_j=
-\int_{-h}^{+h} dx_2 T_{11} 
\label{surf1}
\ee
The spatial gradients parallel to the crack direction ($\partial_1\psi^k$)
give vanishingly small contributions
in the limit $h/\xi\rightarrow +\infty$ and
$R/\xi\rightarrow +\infty$ with $h/R\rightarrow 0$.
Thus, the integrand on the r.h.s of Eq.~(\ref{surf1}) reduces to the energy
of a 1d crack which, as recalled in Appendix \ref{kkl.app} 
(Eq.~(\ref{eexpmp2}))
is itself
independent of the strain and can be identified to twice the interface energy
$\gamma$
\be
F_1^{(coh)}=-\int_{-h}^{+h} dx_2\, \mathcal{E}(\phi,\partial_2\phi,\partial_2 u_2)
=-2\gamma
\label{surf}
\ee
This yields the expected result that
cohesive forces along the crack direction
exert a force opposite to the crack extension force
with a magnitude equal to twice the surface energy. 

One similarly obtains
for $i=2$, in the same limit $\xi \ll h \ll R$,
the other component $F_2^{(coh)}$ of the force
perpendicular to the crack direction
\be
F_2^{(coh)}=\int_{B\rightarrow A} ds \,T_{2j}\,n_j=
-\int_{-h}^{+h} dx_2 T_{21}=\int_{-h}^{+h} dx_2 
\frac{\partial\mathcal{E}}{\partial 
\partial_1 \psi_{\alpha}}\partial_2 \psi_{\alpha}=
\int_{-h}^{+h} dx_2
\frac{\partial\mathcal{E}_{pf}}{\partial
\partial_1 \phi}\partial_2 \phi
\label{herring1}
\ee
The last equality comes from the fact that the only considered anisotropy is
in the phase field part $\mathcal{E}_{pf}$ of the energy density and that,
as above, gradients parallel
to the crack direction give negligible contributions far behind the crack tip.
$F_2^{(coh)}$ can be expressed as the angular derivative of the surface
energy at the crack tip direction $\theta=0$. For a material broken along
a line lying
at a direction $\theta$ with the x-axis, the displacement and phase fields
only depend on the normal coordinate 
$\eta=-x_1 \sin (\theta)+ x_2 \cos(\theta)$. The local energy density
$\mathcal{E}[\phi, \partial_1 \phi,
\partial_2\phi, \partial_{\eta} u_{\eta}]$ is
therefore equal to 
$\mathcal{E}[\phi,-\sin(\theta) \partial_{\eta} \phi,\cos(\theta)
\partial_{\eta}\phi, \partial_{\eta} u_{\eta}]$. The allied surface energy reads
\be
2 \gamma(\theta)=\int_{-\infty}^{+\infty} d\eta\ \mathcal{E}[-
\sin(\theta)\partial_{\eta} \phi,\cos(\theta)
\partial_{\eta}\phi,\partial_{\eta} u_{\eta}]
\label{gpt}
\ee
Differentiation with respect to $\theta$ brings on the r.h.s. of Eq.~(\ref{gpt})
terms coming from the explicit dependence of the integrand on $\theta$
as well as terms coming from the implicit dependence of the fields on the
breaking angle (for an anisotropic material). However, the contribution of
the implicit terms vanishes since  for any given angle the fields minimize
the total energy and no field variation leads to an energy change 
at linear order. Therefore, one obtains
\be
2 \frac{d}{d\theta}\gamma |_{\theta=0}=\int_{-\infty}^{+\infty} dx_2 
\frac{\partial \mathcal{E}}{\partial \partial_1 
\phi}(-\partial_2 \phi)=\int_{-\infty}^{+\infty} dx_2
\frac{\partial \mathcal{E}_{pf}}{\partial \partial_1
\phi}(-\partial_2 \phi)
\label{herring2}
\ee
since $\eta$ reduces to $x_2$ for $\theta=0$. Comparison of 
Eq.~(\ref{herring1}) and (\ref{herring2}) shows that
\be
F_2^{(coh)}=-2 \frac{d}{d\theta}\gamma |_{\theta=0}
\label{f2gpt}
\ee
as announced.

Of course, the relation (\ref{f2gpt}) can also be checked by direct 
computation for any  explicit form of the phase field energy.
For instance, in the simple case of Eq.~(\ref{enani.def}), one obtains
from Eq.~(\ref{herring1}) 
\be
F_2^{(coh)}=\int dx_2 \frac{\partial \mathcal{E}_{pf}}{\partial \partial_1
\phi} \partial_2 \phi =\frac{\epsilon}{2} \int dx_2 \kappa (\partial_2 \phi)^2=
\epsilon \gamma_0
\label{f2expl}
\ee
where, for the last equality, it should be noted (see Appendix \ref{kkl.app})
 that the second integral in 
Eq.~(\ref{f2expl}) 
 is equal to the energy (by unit length) of the cracked material
which is itself equal to 2$\gamma_0$. The result of Eq.~(\ref{f2expl})
indeed agrees with Eq.~(\ref{f2gpt}),
$\epsilon \gamma_0=-2\gamma_{\theta}(0)$, since the interface energy
in the direction $\theta$ is given by Eq.~(\ref{stani2}).

The force $F_2^{(coh)}$  is the direct analog of the Herring torque
$\gamma_\theta=d\gamma/d\theta$ on
grain boundaries \citep[p. 143]{Her1951}. This torque tends
to turn the crack into a direction that minimizes the
surface energy. 

\subsubsection{ Dissipative forces}

The last term in Eq.~(\ref{line}) gives the two components
of the dissipative force
\be
F^{(dis)}_i=v \chi^{-1}\int_{-\infty}^{+\infty}\!\int_{-\infty}^{+\infty}
dx_1dx_2
~\partial_1\phi\partial_i\phi, \label{dissforces}
\ee
The limit where the disk area $\Omega$ tends to infinity has been taken
since the integrand vanishes
outside the process zone.
In contrast to the configurational and cohesive forces,
the  dissipative force clearly depends on the detail of the underlying 
diffuse interface model.

\subsubsection{Force balance and anisotropic generalization of the principle 
of local symmetry}
Substituting the results
of Eqs.~(\ref{f21}) to (\ref{dissforces})
into Eq.~(\ref{line}), the two conditions of Eq.~(\ref{line})
can be rewritten in the compact form
\begin{eqnarray}
F_1&=&G-G_c-F_1^{(dis)}=0,\label{x1}\\
F_2&=&
G_\theta(0)
 -G_{c\theta}(0)
-F_2^{(dis)}=0,\label{x2}
\end{eqnarray}
where we have used the fact
that $G_{c\theta}=2\gamma_{\theta}$.
Eq.~(\ref{x1}) together with Eq.~(\ref{dissforces})  
 predicts the
crack speed 
for $G$ close to $G_c$ 
\be
v\approx \frac{\chi}{\int\!\!\!\int dx_1dx_2 (\partial_1\phi_0)^2}(G-G_c)
\label{v.eq}
\ee
where $\phi_0$ is the phase-field
profile for a stationary crack \citep{KarLob2004}, and thus the integral
in the denominator above is just a constant of order unity.
Eq.~(\ref{x2}), in turn, predicts the crack path by imposing
$K_2$ at the crack tip,
\be
K_2=-\left(
G_{c\theta}(0)
+F_2^{(dis)}\right)/(2\alpha K_1).
\label{k2}
\ee
The component $F_2^{(dis)}$ of the dissipative force vanishes
with the crack velocity in the quasitatic limit. So, in this limit,
the microscopic details of the process zone do not play a role and
the crack direction is uniquely determined by the directional
anisotropy of the material through the simplified condition
\be
K_2=-G_{c\theta}(0)/(2\alpha K_1).
\label{k2simp}
\ee
Eq.~(\ref{k2simp}) replaces the principle of local symmetry for a material
with an anisotropic surface tension energy. It
reduces of course to the principle of local symmetry
in an isotropic material,
since then $G_{c\theta}$ vanishes. One can also note that quite remarkably,
Eq.~(\ref{k2simp}) only contains macroscopically defined parameters and is 
independent of the detailed physics of the process zone.

Outside the quasistatic limit, $K_2=0$ should continue to hold for
an isotropic material since $F_2^{(dis)}$
vanishes even for a finite crack speed.
The latter follows from the symmetry of the inner
phase-field solution for a propagating crack with $K_2=0$,
$\phi(x_1,x_2)=\phi(x_1,-x_2)$, which implies that
the product
$\partial_1\phi\partial_2\phi$ in Eq.~(\ref{dissforces}) is
anti-symmetric and that the spatial integral of this product 
vanishes. In an anisotropic material, however,
$\phi$ is generally not symmetrical about the local crack axis
and $F_2^{(dis)}$ should generally be non-zero. The crack direction should then
become dependent on the details 
of the energy dissipation in the process zone. 

A small velocity expression
for the dissipative force perpendicular to the crack axis can be
obtained by considering the phase-field
profile that corresponds to a stationary Griffith crack in
an anisotropic material. Here, $\phi_0^A$, 
is uniquely defined as the stationary phase-field profile that exists
for a unique pair of values of $K_1$ and $K_2$ that satisfy the conditions
of equilibrium parallel and perpendicular to the crack axis,  
$\alpha(K_1^2+K_2^2)=G_c(0)$ and $-2\alpha K_1K_2=G_{c\theta}(0)$,
respectively. For small velocity, Eq. (\ref{dissforces}) must therefore
reduce to $F_2^{(dis)}=v\chi^{-1} I(0)$ where the integral
\begin{equation}
I(0)\equiv \int dx_1dx_2~\partial_1\phi_0^A\partial_2\phi_0^A
\end{equation}
is a dimensionless constant that, like $G_c$ and $G_{c\theta}$, depends 
generally on the local orientation of the crack with respect to some fixed reference axis
chosen here as $\theta=0$. 
For small velocity, Eq. (\ref{k2}) therefore becomes
\be
K_2=-\left(
G_{c\theta}(0)
+v\chi^{-1}I(0\right)/(2\alpha K_1),
\ee
where $I(0)$ vanishes in the isotropic limit since $\phi_0^A$ approaches $\phi_0$ and hence becomes
symmetrical about the crack axis in that limit.


\subsection{Comparison with the maximum energy release rate criterion}
The principle of local symmetry and the maximum energy release rate criterion
gives slightly different results in general, for instance for the prediction
of the finite angle of a kink extension at the tip of a crack. The two criteria
coincide however for smooth cracks. It is interesting to note that it is also
true for the present anisotropic generalisation (Eq.~(\ref{k2simp})) of
the principle of local symmetry.
One way to generalize the maximum energy release rate criterion
for anisotropic material 
is to require the crack growth to take place in the direction that maximizes
$\tilde{G}(\theta)-2\gamma(\theta)$ \citep{lebprivcom}
where as before $\tilde{G}(\theta)$ is the energy release rate for an 
infinitesimal extension at the crack tip at an angle $\theta$ (where as
before $\theta=0$ is the direction of the unextended crack).  
For a smooth crack, the condition that this quantity be maximal in the 
crack direction yields
\be
\frac{d}{d\theta}[\tilde{G}(\theta)-2\gamma(\theta)]_{\theta=0}=0
\ee
With the help of Eq.~(\ref{dgdcal}), this is seen to be identical to Eq.~(\ref{k2simp}) as stated.

\subsection{Crystalline materials}
Our results have interesting implications
for crack propagation in crystalline materials. Basic experimental studies have demonstrated the existence of both ``cleavage 
cracks'', which are cracks that propagate along low energy crystal 
planes, such as $\{111\}$  \citep{Hauetal1999}  or $\{110\}$  \citep{Deeetal2003} in silicon, and
smooth cracks \citep{Deeetal2003} that resemble qualitatively the cracks seen in isotropic materials.
While the propensity for crack propagation along
cleavage planes in crystalline materials is to be expected energetically, 
the observation of smooth cracks in those same materials is perhaps less intuitive.
Theoretical attempts have been made to understand when cracks will 
cleave crystals using both energetic arguments and lattice simulations \citep{Deeetal2003,Mar2004}.
However, a consistent theoretical picture has not yet emerged. 

The crack propagation law  Eq.~(\ref{k2simp}) provides an explicit prediction
of when a crack will propagate along a cleavage plane, or smoothly in other
directions.  Restricting our discussion to
two dimensions for simplicity, the surface energy in a crystalline material 
is expected to show a cusp behavior  
\be
\gamma(\theta)=\gamma_0(1+\delta|\theta|+\dots),
\ee
near a cleavage plane (and more generally near sets of equivalent low
energy crystal planes imposed by symmetry), 
where $\theta$ measures the angle of the surface  
away from this plane, and to be a smooth differentiable function of $\theta$ for other orientations.
In terms of the physical picture outlined in Section 2, this cusp behavior
implies the presence of a finite Herring torque on any small extension of a crack at an
infinitesimal angle away from a cleavage plane. Therefore, a crack will be essentially trapped
along a cleavage plane until the Eshelby configurational torque is large
enough to tilt the crack away from this plane.
Restated in terms of the propagation law,
Eq.~(\ref{k2simp}) can be obeyed for small non-zero angles
only when $|K_2|$  exceeds a threshold $K_2^{(c)}$ with
\be
K_2^{(c)}=\frac{E \gamma_0\delta}{(1-\nu^2)K_1}, \mathrm{for} \ G\approx G_c. \label{K2threshold}
\ee
Consequently, $|K_2|$ should exceed $K_2^{(c)}$
for a cleavage crack to change direction. Eq.~(\ref{k2simp}) also
implies that a crack
will propagate smoothly for other orientations away from cleavage planes where the  
surface $\gamma$-plot is smooth.  

One interesting prospect to test this prediction is to
examine its consequences for thermal fracture in crystalline materials,
where quasistatic oscillatory cracks have been studied under
well-controlled experimental conditions. In particular, experiments 
have revealed that the onset of crack oscillations is markedly different in
anisotropic and isotropic materials. In crystalline silicon wafers that
cleave preferentially $\{110\}$ planes, the onset of crack oscillations is delayed 
in comparison to an isotropic material and is accompanied by a discontinuous jump
in oscillation amplitude consistent with a subcritical bifurcation \citep{Deeetal2003}.
In contrast, the onset of thermal crack oscillations in isotropic material has
been shown to be supercritical in a recent phase-field modeling study,
consistent with earlier experimental observations in glass (see \citep{Coretal2008} and earlier
references to the experimental literature therein). The existence of a finite threshold Eq. (\ref{K2threshold})
to escape a cleavage crack precludes a smooth transition to crack oscillations around a
cleavage plane. One would therefore expect a subcritical bifurcation for the onset of crack
oscillations if Eq.~(\ref{k2simp}) is used in conjunction with a typical
$\gamma$-plot for a crystalline material that exhibits cusps. However, a detailed
study is clearly needed to validate this expectation and to make contact
quantitatively with experiments.

\section{Motion under pure antiplane shear}
\label{mo3.sec}
As recalled in the introduction, the principle of local symmetry was first
proposed in \citep{BarChe1961} for crack motion under 
pure antiplane shear. This particular case does not seem to have attracted
much interest subsequently, presumably because rotation of the crack front
is observed  and fractures under  mixed mode I-III loading
are found to be unstable in three dimensional isotropic materials
\citep{Sommer69}. The  criterion of motion under antiplane shear
could nonetheless have some importance for the development of the
tridimensional instability. It is also
conceivable that 2D motion under pure mode III loading could be 
effectively realized in an appropriate anisotropic material, like for
instance a thin layer of sintered glass beads. We therefore find it interesting
to briefly examine  this criterion for
motion under pure antiplane shear with the formalism developed in
the previous sections. 

Since in a pure mode III motion, the displacement
field reduces to its third component $u_3$ that is a purely scalar Laplacian
field, the diverging stress distribution near the tip Eq. (\ref{sigma3}) is always symmetrical
and produces no configurational force perpendicular to the crack axis. Consequently, for
propagation in an anisotropic material, the propagation law reduces simply to
the condition that the Herring torque vanishes, $\gamma_{\theta}=0$.
This condition implies that in the limit of vanishing velocity,
a quasistatic crack propagates in a direction that corresponds
to a local minimum of the surface energy; it can be argued that
propagation in a direction of maximal $\gamma$ is unstable because the configurational torque
amplifies small departures from this direction. 
Furthermore, for finite velocity, the dissipative force perpendicular to the crack axis, $F_2^{(dis)}$,
also vanishes since the phase-field profile must be symmetrical
about the crack axis for a direction where $\gamma_{\theta}=0$.

For propagation in an isotropic material, the situation is more subtle than for
the mode I/II case. The evaluation of the contour integral that
is the direct analog for mode III of the r.h.s. of Eq. (\ref{scysimp})
(or equivalently Eq. (\ref{fconf})),
gives only a non-vanishing force perpendicular to the crack axis if the subdominant 
antisymmetrical contribution of the stress distribution (i.e., the second term on the r.h.s. of Eq. (\ref{sigma3})
 is included. This force is proportional to $K_3A_2\sqrt{R}$, where $R$
is the radius of the integration contour around the crack tip, where the square-root
behavior follows from dimensional analysis. This force vanishes if $A_2=0$, thereby
suggesting that the original formulation of the principle of local symmetry for mode III
might be applicable. However, our inner-outer matching procedure used to compute this force is predicated
on choosing $R$ much larger than the scale of the process
zone but vanishingly small on the outer scale of the system
size set by material boundaries. Therefore, the magnitude of this
force is left undetermined in the present analysis. Further work 
is therefore needed to determine if  the inner and outer scales can be clearly
separated for pure antiplane shear and if 
$A_2=0$ can rigorously serve as a local
condition to predict crack paths in isotropic material.

Additional insight into this question can be gained by repeating the analysis
of Section 2 for a small extension $\delta l$ of a mode III crack.
The analogous expression for the stress intensity factor at the
tip of the extended crack is
$K_3^*=K_3-b \mu A_2 \sqrt{\delta l}\delta \theta$ to linear
order in $\delta \theta$ \citep{Sih1965} where $b$ is a numerical
constant, and hence $G_\theta(0)\sim K_3A_2 \sqrt{\delta l}$.
One important difference with plane loading
is the square-root dependence of $G_\theta(0)$
on the crack extension length, which is also
reflected in the $\sqrt{R}$ dependence of the integral
just mentioned above, which yields the configurational force perpendicular to
the crack tip for mode III. Since the only
natural cut off for the crack extension length on the outer scale of the system is the
size $\xi$ of the process zone, this result seems to imply
that  $G_{\theta}(0)\sim K_3A_2 \sqrt{\xi}$
up to a numerical prefactor. It also yields the local symmetry condition $A_2=0$
in the isotropic limit, where the symmetry
of the phase-field profile
makes the dissipative force $F_2^{(dis)}$ vanish. 


\section{Numerical simulations and tests}
\label{num.sec}

We focus here on testing the relation (\ref{k2simp}) between $K_2$ at
a crack tip and the derivative of the interface energy, in the case of
plane strain. We numerically compute
the  extension of a preexisting straight crack as described by
the phase-field equation
(\ref{pfe}) and (\ref{stani2}). For a pure mode I loading 
of the preexisting crack and an anisotropic surface energy, 
Eq.~(\ref{k2simp})  predicts that
the growth of a kinked extension takes place at an angle $\theta$ such  that
$K_2$ is adequate at the growing tip. More explicitly,
on the one hand,
Eq.~(\ref{ktilde}) gives
$\tilde{K}_2$ at kink tip as
\be
\tilde{K}_2= F_{21}(\theta) K_1\simeq K_1\frac{\theta}{2}
\ee
where the second equality is valid for small angles.
Therefore, one obtains for small angles,
\be
-2\alpha \tilde{K}_1 \tilde{K}_2\simeq \alpha K_1^2\frac{\theta}{2}
\label{num1}
\ee
since $\tilde{K}_1$ is equal to $K_1$ at dominant order in $\theta$ 
(Eq.~(\ref{ktilde}).
On the other hand, the surface energy (\ref{stani2}) gives 
\be
-2\gamma_{\theta}(0)=\epsilon\gamma_0
\label{num2}
\ee
Eq.~(\ref{k2simp}), which translates in the equality of the l.h.s. of
Eq.(\ref{num1}) and (\ref{num2}), simplifies for $G$ close to 
Griffith threshold when $\alpha K_1^2\simeq 2 \gamma_0$.
Then, it simply gives for the initial angle $\theta$ of
the kink crack
\be
\theta=\frac{\epsilon}{2}
\label{pred}
\ee
which is strictly valid for $\epsilon\ll 1$ in the limit
$G\rightarrow G_c$.

This prediction was tested numerically.
Eq.~(\ref{pfe}) was solved by using
an Euler explicit scheme to integrate
the phase-field evolution and a successive
over relaxation (SOR) method to calculate the quasi-static
displacement fields $u_1$ and $u_2$
at each time step. We used as initial
condition a straight horizontal crack of length $2W$
centered in a strip of length $4W$ horizontally and $2W$ vertically,
with fixed values of $u_1$ and $u_2$ on the strip boundaries
that correspond to the  singular stress fields defined by Eq.~(\ref{km}) for
prescribed values of $K_1$ and $K_2$.
We used $\lambda/\mu=1$ [$\alpha=3/(8\mu)$], ${\mathcal E}_c/\mu=1/2$, a grid spacing
$\Delta x_1=\Delta x_2=0.1\xi$, and $W=50 \xi$,
where the process zone size
$\xi\equiv \sqrt{\kappa/(2{\mathcal E}_c)}$.
We checked that the results were
independent of width and grid spacing.

\begin{figure}[t]
\begin{center}
\includegraphics[width=8cm]{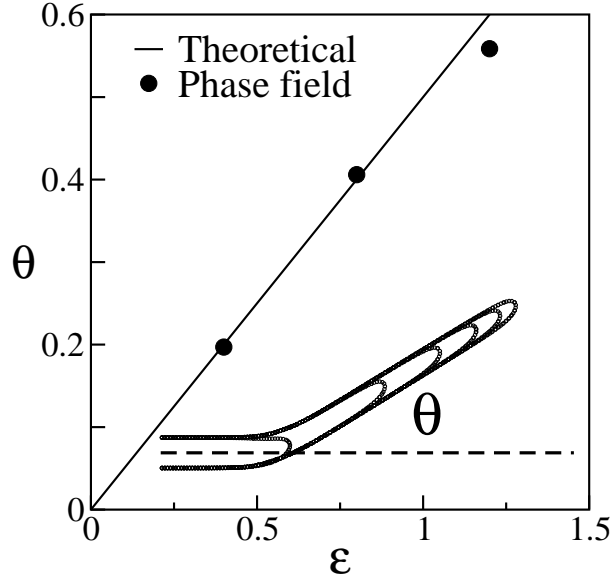}
\end{center}
\caption{Kink angle $\theta$ versus
surface energy anisotropy $\epsilon$ for plane strain
predicted as $\theta=\epsilon/2$
and extracted from phase-field simulations (filled circles)
for $G/G_c\approx 1.1$.
Inset: phase-field simulation showing $\phi=1/2$ contours
equally spaced in time
for $\epsilon=1.2$.}
\label{compare}
\end{figure}

We  first verified that, in the isotropic limit, the kink angle
was well predicted by the local symmetry
condition $K_2^*=0$
, which implies
that $\theta\approx -2K_2/K_1$.
Then, for the anisotropic
case, we chose $K_2=0$ and $G$ just slightly above
$G_c$.
The results for the kink angles observed for several simulations with
different magnitudes of the surface energy anisotropy $\epsilon$
are shown in Fig.~\ref{compare}.
The prediction (\ref{pred})  is seen to be in good quantitative agreement
with the results of the phase-field
simulations.

We have also tested our prediction for pure mode III cracks.
We used the same phase-field model and 
anisotropy form of $\gamma(\theta)$ as for the plane strain case,
albeit with the strain energy corresponding to pure antiplane shear 
\begin{equation}
\mathcal{E}_{strain}=\frac{\mu}{2}\left[(\partial_1u_3)^2+(\partial_2u_3)^2\right].
\end{equation}
We used as initial
condition a straight horizontal crack of length $3W$
centered in a strip of length $6W$ horizontally (along the $x_1$ axis) and $2W$ vertically
together with ${\mathcal E}_c/\mu=1/2$, the anisotropy $\epsilon=1.8$, a grid spacing
$\Delta x_1=\Delta x_2=\xi/6$,  a half strip width $W=50 \xi$,  and a 
fixed displacement $u_3= 11.313~\xi$ on the $x_2=\pm W$ boundaries
corresponding to a crack slightly above the Griffith threshold,
where as before the process zone size
$\xi\equiv \sqrt{\kappa/(2{\mathcal E}_c)}$.

\begin{figure}[t]
\center
\includegraphics[width=8cm]{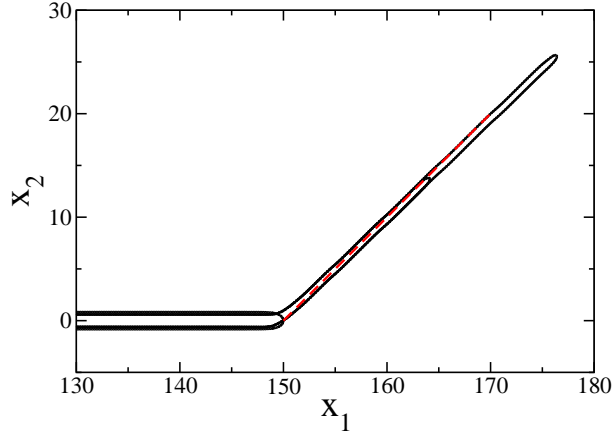}
\caption{Phase-field simulation for pure antiplane shear
for $\epsilon=1.8$ ($\phi=1/2$ contours are equally
spaced in time). The dashed line is a guide to
the eye for the 45$^o$ kink angle predicted by
the vanishing torque condition $\gamma_\theta=0$.} 
\label{mode3kink}
\end{figure}

The results of this simulation shown in Fig. \ref{mode3kink} confirm
that a crack centered initially in the strip with its axis parallel
to the $\theta=0$ direction, kinks at a
45$^o$ angle ($\theta=\pi/4$) that is consistent with the
analytical prediction $\gamma_\theta=0$ for mode III
in an anisotropic material. 

Fracture in the phase-field model that we have considered here is a reversible 
process in the sense that cracks can (and do)
heal when stresses are removed. This can also be observed in some experiments 
under very clean conditions when no alterations of exposed surfaces follow breaking.
Nonetheless, this
is sometimes considered a troublesome feature since it does not occur in more
usual conditions. One could think of introducing irreversibility in a "physical
way" by adding another field to mimick surface oxydation. In a simpler but
more
{\em ad-hoc} fashion, one can only accept evolutions that
decrease the value of the phase field.
To assess the importance of reversibility on our results, the numerical 
simulations above for mode I/II and mode III were redone with this second
scheme ((i.e. taking Eq.~(\ref{pfe}) as written
when $\partial_t \phi <0$ and otherwise replacing it by $\partial_t \phi=0$).
Reassuringly, the results are essentially identical to those
plotted in Fig.~\ref{compare} where the $\phi=1/2$ phase-field contours
superimpose perfectly for the two sets of simulations with and without
reversible dynamics. This insensitivity of the results to the introduction of
irreversibility does not appear surprising since our derivation of crack
propagation laws for modes I/II and III rely on the {\it existence} of propagating
solutions for which $\partial_t \phi<0$. 

\section{Conclusion}
\label{con.sec}
We have analyzed here the laws of quasistatic crack tip motion within the
phase-field framework. The analysis provides a derivation of the principle
of local symmetry and of its generalization to anisotropic materials. It also
underlines the role of the configurational
force perpendicular to the crack tip direction. The results can be
interpreted physically  as a simple force balance condition.
The variational character of the phase-field equations of motion played
an important role in the derivation of the equations for the crack tip. It
directly
allowed us in section \ref{esh.sec} to define a generalization of Eshelby tensor
that includes the phase-field and short-scale physics while keeping its
divergenceless property. Its role in the derivation of section \ref{sol.sec}
 may be less central from a conceptual point of view but the
self-adjointness of the linear operator  allowed
the obtention of explicit formulas. In any case, it should be noted that direct
link to energy considerations {\em \`a la} Griffith require a variational model.

Several questions appear worth of further investigations. First, even if 
pure mode III cracks are not realized under most experimental conditions,
the applicability of the principle of local symmetry ($A_2=0$) remains
to be established on a firmer footing in the isotropic limit where the force
perpendicular to the crack tip seems to depend on an arbitrary cut-off on the
scale of the process zone. Second, the 
crack propagation laws have been derived for a gradient
formulation of the phase-field dynamics where failure is reversible.
Even though our numerical simulations indicate that crack paths are not
altered by a simple ad-hoc introduction of irreversibility where the phase-field
can only decrease, the role of irreversibility 
is worth investigating more fully.

The phase field energy is certainly another aspect that would benefit from
further refinement. As written in Eq.~(\ref{eliso.def},\ref{pfe}), it does not
distinguish between compressive and extensive strains which is quite an 
unphysical feature. Note that this property is also shared by
 the variational formulation
advocated by  \citet{Fra98} 
(illustrated in numerical simulations of  ~\citep[Fig.~4]{Bou00}
in which it is referred to as sample interprenetration).
Some remedies have been proposed \citep{HenLev2004} that break
the variational character of the equation of motion and
do not appear
entirely satisfactory. The development of more physically motivated and material adapted energies
certainly appear as an interesting future endeavor.

The extension of the present analysis 
to three dimensions where
fracture paths are geometrically more complex
is another  important future direction. Numerical simulations
and preliminary analysis addressing this question will be reported elsewhere
\citep{ponskarma07}.

Finally, we hope that the results reported here will contribute to
stimulate further
experimental investigations of quasistatic crack motion of cracks in 
anisotropic media.



We thank M. Adda-Bedia, J. B. Leblond, A. Chambolle, G. Francfort
and J.~J. Marigo for
valuable discussions and instructive comments. A.K. acknowledges the support
of DOE Grant No.  DE-FG02-07ER46400 and the hospitality of the Ecole Normale
Sup\'erieure in Paris where part of this work was completed.

\appendix
\section{The KKL phase-field model in one dimension}
\label{kkl.app}

\begin{figure}
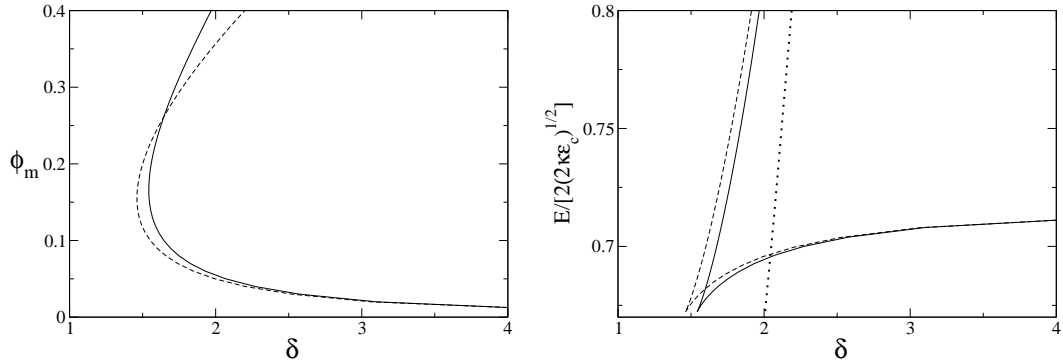

\begin{tabular}{ll}
\bf A & \bf B \\
\includegraphics[scale=0.27]{phimvsdellpap.eps} &
\includegraphics[scale=0.27]{envsdellpap2.eps}
\end{tabular}

\caption{({\bf A}) Minimum value $\phi_m$ of the phase field {\em vs.}
adimensionned displacement $\delta=\Delta (\mu/\kappa)^{1/2}$ for a strip
of adimensionned half-width $\ell=L (2 \mathcal{E}_c/\kappa)^{1/2}=3$ 
(solid line) and in the limit of a large strip (dashed line), 
as given by Eq.~ (\ref{eqd4}) with $\ell=3$.
For a given displacement $\delta>\delta_c$, there are two values of
$\phi_m$ corresponding to two stationary solutions with the smallest value of 
$\phi_m$ for the stable one.
({\bf B}) The energies $E$ vs. $\delta$
for the two branches of solutions for the same band of
$\ell=3$ (solid line), the lower branch corresponding to the stable fractured 
state. The asymptotic expression for a large band is also plotted using
Eq.~(\ref{enas}) with $\ell=3$ (dashed line).
The two branches
meet at a cusp, the generic behavior at a saddle-node bifurcation. The 
corresponding adimensionned energy 
$\delta^2/(2\ell)$ of
the (half) elastically-stretched band is also plotted (dotted line). It becomes larger
than the energy of the stable fractured solution at  Griffith threshold.
}
\label{app.fig}
\end{figure}

In this appendix, we recall the analysis of the KKL phase-field model in one 
dimension, that is the snap-back of a stretched elastic band,
as described in \citep{KKL2001}. In particular, the energy of the fractured state 
Eq.~(\ref{eexpmp2}) provides the expression of the interface energy 
given in the main text.
We also  show how the fractured solution appears in one dimension in this model.
 For an elastic band of size $2 L$,  
the elastically stretched state is the only allowed state
 when the total strain $2 \Delta$ is low 
enough. Above a critical total 
strain $2 \Delta_c$ two other non trivial solutions appear via a 
saddle-node bifurcation, one being dynamically stable and the other being
 unstable. At the bifurcation, both solutions have a higher energy than the
elastically streched state. However, 
the dynamically stable solution becomes energetically favored as compared to
the 
elastically streched state when the total strain becomes higher than 
$2 \Delta_G\, (> 2 \Delta_c)$, which corresponds to  Griffith threshold in the model. This scenario is illustrated by numerical solution in Fig.~\ref{app.fig}. 
The unstable solution corresponds to the energy barrier (the Eyring state)
that has to be overcome to create the fractured state and it provides the
corresponding activation energy.

For a one-dimensional band, the KKL energy reads,
\begin{equation}
E=\int_{-L}^{+L} dy\left\{\frac{\kappa}{2}(\partial_y\phi)^2+
g(\phi)[(\frac{\lambda}{2}+\mu)(\partial_y u)^2-\mathcal{E}_c]+\mathcal{E}_c \right\}
\label{e1d}
\end{equation}
with the function $g$ monotonically increasing from $g(0)=0$ in the fully broken
state to $g(1)=1$ in the intact material with also $g'(1)=0$ to recover linear
elasticity.
Steady state solutions obey the equilibrium equation obtained by variation 
of Eq.~(\ref{e1d}),
\begin{eqnarray}
\kappa \partial_{yy}\phi &=&g'(\phi)
[(\frac{\lambda}{2}+\mu)(\partial_y u)^2-\mathcal{E}_c]\label{eqph}\\
\partial_{y}[g(\phi)\partial_y u]&=& 0\label{equ}
\end{eqnarray}
with the boundary conditions $u(\pm L)=\pm \Delta,\ \phi(\pm L)=1$. 
The elastically stretched band $\phi=1,\ u(y)=y\Delta/L$ is always a solution
of Eqs.~(\ref{eqph}, \ref{equ}) and its energy is equal to the usual purely
elastic one 
\begin{equation}
E=(\lambda+2\mu)\Delta^2/L.
\label{eqel}
\end{equation}

In order to analyze the existence of other less obvious solutions of
Eqs.~(\ref{eqph},\ref{equ}), it is useful to note that Eq.~(\ref{equ})
can be integrated once to obtain
\begin{equation}
\partial_y u=\sqrt{\frac{\mathcal{E}_c}{\lambda/2+\mu}}\ \frac{c}{g(\phi)}
\label{uph}
\end{equation}
with $c$ a constant yet to be determined.
Eq.~(\ref{uph}) allows the elimination of the strain field from the phase-field equation
(\ref{eqph}) which then reads
\begin{equation}
\kappa \partial_{yy}\phi =\mathcal{E}_c g'(\phi)\left[ \frac{c^2}{g^2(\phi)}
- 1 \right]
\label{eqph2}
\end{equation}

In a usual way, it is helpful to consider y as a fictitious time and to
think of
Eq.~(\ref{eqph2}) as describing the motion of a point 
particle in the effective potential
\begin{equation}
V_{eff}(\phi)=\frac{c^2}{g(\phi)}+ g(\phi)
\end{equation}
With this analogy, a non-trivial solution of Eq.~(\ref{eqph2}) corresponds to
a particle that starts  at ``time'' $y=-L$ from $\phi=1$ with a negative velocity
$\partial_y\phi < 0$, to reach a minimum $\phi=\phi_m$ at $y=0$ where the 
``velocity'' $\partial_y\phi$ vanishes; 
 from this turning point it then follows the time-reversed motion and comes back
to $\phi=1$
at $y=+L$. The integrability of this one-dimensional motion gives the
conservation law
\begin{equation}
\frac{\kappa}{\mathcal{E}_c} (\partial_y\phi)^2 + V_{eff} (\phi) = 
V_{eff} (\phi_m)
\label{enmot}
\end{equation}
This allows us 
to express the energy (Eq.~(\ref{e1d}) of the corresponding non-trivial solution as
\begin{equation}
E= \sqrt{2\kappa\mathcal{E}_c}\int_{\phi_m}^1\frac{d\phi}{\sqrt{V_{eff}(\phi_m)-
V_{eff}(\phi)}} [1+ V_{eff}(\phi_m)- 2 g(\phi)]
\label{eexpmp}
\end{equation}

Two constraints determine the two unknown constants $c$ and $\phi_m$ as a 
function of the dimensionless strip width $\ell$ and dimensionless total strain
$\delta$. First,
the particle motion should take
a total time $2 L$ with
\begin{equation}
\ell=L\sqrt{2 \mathcal{E}_c/\kappa}
=\int_{\phi_m}^1\frac{d\phi}{\sqrt{V_{eff}(\phi_m)-V_{eff}(\phi)}}
\label{eql}
\end{equation}
Second, the overall integrated strain should equal the imposed total strain
\begin{equation}
\delta=\Delta\sqrt{(\lambda+2 \mu)/\kappa}
=c \int_{\phi_m}^1\frac{d\phi}{g(\phi) \sqrt{V_{eff}(\phi_m)-V_{eff}(\phi)}}
\label{eqd}
\end{equation}
We analyze more specifically
 the case of a macroscopic strip of width $\ell\gg 1$.

We find it convenient to first consider the dependence of 
Eq.~(\ref{eql}) on $c$. Since the sum of kinetic energy 
[$\kappa/(2\mathcal{E}_c) (\partial_y\phi)^2)]$ and potential
energy [$V_{eff}(\phi)$] is constant along the particle trajectory 
(Eq.~(\ref{enmot}), 
the initial potential energy is always lower than the 
potential energy at the return point where the kinetic energy vanishes),
  $V_{eff}(1)<V_{eff}(\phi_m)$. For a given $\phi_m$, the time $L$ 
spent by the particle during its motion increases as its  initial velocity 
$\vert{\partial_y\phi}\vert$ decreases. It is thus maximal in the limit where 
$V_{eff}(1)=1+c^2$ tends towards $V_{eff}(\phi_m) =
c^2/g(\phi_m)+ g(\phi_m)$, that is in the limit $c^2\rightarrow g(\phi_m)$. 
The time spent on the trajectory diverges logarithmically when $c^2$ 
approaches $g(\phi_m)$ (since $g'$ has a double zero at $\phi=1$). Thus, for
$\ell\gg 1$, $c^2$  is exponentially close to $g(\phi_m)$.

We consider now the total strain constraint Eq.~(\ref{eqd})
and the determination of $\phi_m$. It is helpful to rewrite Eq.~(\ref{eqd})
 using 
Eq.~(\ref{eql}) as
\begin{equation}
\delta
=c \ell + c \int_{\phi_m}^1 d\phi 
\frac{1-g(\phi)}{g(\phi) \sqrt{V_{eff}(\phi_m)-V_{eff}(\phi)}}
\label{eqd2}
\end{equation}
Under this form, for large $\ell$, as $c^2\rightarrow g(\phi_m)$
 the integral
in Eq.~(\ref{eqd2}) converges and  $c^2$ can be replaced by $g(\phi_m)$ 
with an exponentially small error. Thus, one obtains
\begin{equation}
\delta \simeq
g(\phi_m)^{1/2}\left\{ \ell + 
\int_{\phi_m}^1 d\phi \sqrt{\frac{1-g(\phi)}{g(\phi) [g(\phi)-
g(\phi_m)]}}\right\}
\label{eqd3}
\end{equation}
The existence of solutions with $\delta \ll \ell$ 
imposes $g(\phi_m)\ll 1$ (since
the integral term in Eq.~(\ref{eqd3}) is clearly positive). When $g$
behaves as $g(\phi)\sim a \phi^{\sigma}$ for  small
Eq.~(\ref{eqd3})  reduces to
\begin{equation}
\delta \simeq
\sqrt{a} \phi_m^{\sigma/2} \ell + \frac{C_{\sigma}}{\sqrt{a}}
\phi_m^{1-\sigma/2}
\label{eqd4}
\end{equation}
where the constant $C_{\sigma}$ can be expressed in term of the Euler $B$ 
function as
 $C_{\sigma}=B(1-1/\sigma,1/2)/\sigma$. 
The first term in Eq.~(\ref{eqd4}) represents a contribution to the total 
displacement that is distributed over the whole sample whereas the second one is
a localized contribution coming from the center of the stretched band. 
If one wishes that some solutions can correspond to fractured bands, the second
contribution should dominate the first. It should moreover
 be able to take values much larger than one (so than one can have localized
solutions with $\delta\gg 1$).
 This clearly requires  the  exponent
$1-\sigma/2$ to be negative and therefore that the function $g(\phi)$ be
chosen so that $\sigma>2$, as noted in \citep{KKL2001}. A possible choice, made in the present work
 as in \citep{KKL2001}, is to take 
$\sigma=3$
(in addition to the requirements $g(1)=g'(1)=0$) and
\begin{equation}
g(\phi)=4 \phi^3 -3\phi^4
\label{eqg}
\end{equation}
For this specific choice,
clearly $a=4$ and $C_3=B(2/3,1/2)/3\simeq 0.862\,$.
For $\sigma>2$, the r.h.s. of Eq.~(\ref{eqd4}) has a minimum value $\delta_c$
that is reached for $\phi_m=\phi_c$ with 
\begin{eqnarray}
\phi_{c}^{\sigma-1}&\simeq& C_{\sigma}(\sigma-2)/(a\sigma\ell) \\
\delta_c&\simeq&\sqrt{a}\frac{2(\sigma-1)}{\sigma-2} \ell \phi_{c}^{\sigma/2}\sim
\ell^{\frac{\sigma-2}{2(\sigma-1)}}
\end{eqnarray}
or more simply for our specific choice of $g$ with $\sigma=3$,
$\phi_{c}\simeq 0.268/\sqrt{\ell}$ and $\delta_c \simeq 1.11 \ell^{1/4}$.
For an adimensionned strain $\delta$ below $\delta_c$ no non-trivial
solutions exist. Two coincident solutions appear at $\delta=\delta_c$ which
separate into a stable lower energy solution and an unstable higher energy one
when $\delta>\delta_c$ as shown in Fig.~\ref{app.fig}. The energies of the
two solutions can be explicitly obtained in the limit $\ell\gg 1$.
Eq.~\ref{eexpmp} can be rewritten as
\begin{equation}
\frac{E}{\sqrt{2\kappa\mathcal{E}_c}}=[V_{eff}(\phi_m)-1] \ell +
 \int_{\phi_m}^1\frac{d\phi}{\sqrt{V_{eff}(\phi_m)-
V_{eff}(\phi)}} 2[1-  g(\phi)] 
\label{eexpmp3}
\end{equation}
where we have used the expression (\ref{eql}) for the strip width $\ell$.
For a large $\ell$, $c^2$ can be replaced by $g(\phi_m)$ with an
exponentially small error to obtain
\begin{equation}
\frac{E}{\sqrt{2\kappa\mathcal{E}_c}}= g(\phi_m) \ell +
2 \int_{\phi_m}^1 d\phi [1-  g(\phi)]\sqrt{\frac{g(\phi)}{g(\phi)-g(\phi_m)}}
\label{eexpmp4}
\end{equation}
Finally, in the whole regime of interest where $\delta\ll\ell$, the phase-field
minimum value $\phi_m$ vanishes as a power of $\ell$. With the small
$\phi$ behavior $g(\phi)\sim a \phi^{\sigma}$, Eq.~(\ref{eexpmp4}) simply 
reduces to
\begin{equation}
\frac{E}{\sqrt{2\kappa\mathcal{E}_c}}=
 a \phi_m^{\sigma}\ell
- D_{\sigma}\phi_m +
2 \int_{0}^1 d\phi [1-  g(\phi)]
\label{enas}
\end{equation}
where as above the constant $D_{\sigma}$ can be expressed using Euler B function
($D_{\sigma}=(\sigma-2) C_{\sigma}$) and for $\sigma=3$, $D_3\simeq 0.862$.
The asymptotic form (\ref{enas}) is already reasonnably accurate for $\ell=3$ as shown
in  Fig.~\ref{app.fig}.

As $\delta$ becomes much larger than $\delta_c$, these two solutions correspond
to the dominance of one of the two terms on the l.~h.~s.~ of Eq.~(\ref{eqd3}).

For the stable solution, the
localized contribution to the strain dominates so that
\begin{equation}
\delta \simeq
 \frac{C_{\sigma}}{\sqrt{a}}
\phi_m^{1-\sigma/2}
\label{eqd5}
\end{equation}
Thus, in this parameter regime, $\ell\gg 1$ and $\delta\gg\delta_c\gg 1$,
 $\phi_m$ tends
toward zero. As a welcome consequence,
the energy of the stable solution becomes 
independent of the strain and can be identified with twice the surface energy
$\gamma$
\begin{equation}
E_s= 2 \gamma=2 \sqrt{2\kappa\mathcal{E}_c}\int_0^1 d\phi\sqrt{1-
 g(\phi)}
\label{eexpmp2}
\end{equation}
For our specific choice of $g(\phi)\,$ [Eq.~(\ref{eqg})] the
numerical value of the integral is approximately 0.7165. 
As in Griffith's original theory,
this stable solution becomes energetically favored as compared to the 
elastically stretched band when $E_s$ becomes smaller than the purely
elastic stretching energy [Eq.~(\ref{eqel})],
that is when 
 $\Delta > \Delta_g=\sqrt{[2\gamma/(\lambda+\mu)] L}$ or equivalently
 for $\delta>\delta_g=1.20\sqrt{\ell}.$

For the unstable solution, when $\delta$ becomes much larger than
$\delta_c$, one has simply
\begin{equation}
\delta \simeq
\sqrt{a} \phi_m^{\sigma/2} \ell 
\end{equation}
The corresponding energy is simply (Eqs.~(\ref{enas}),(\ref{eexpmp2})),
\begin{equation}
E_u= 2 \gamma + \sqrt{2\kappa\mathcal{E}_c}\ \frac{\delta^2}{\ell}
\end{equation}
since the term proportional to $\phi_m$ becomes subdominant with respect to the
other two (and tends towards zero) when $\delta$ moves away from $\delta_c$.
In other terms, the energy of the unstable state is simply the energy of the
elastically stretched band plus the energy necessary to create the
the two interfaces, as one could have
intuitively guessed.

Finally, it is interesting to see how the profile of the fracture state
depends on the total strain $\delta$.
The profile of a general solution is
obtained from Eq.~(\ref{eqph2}) as
\begin{equation}
\frac{y}{\xi}=\int_{\phi_m}^{\phi}
\frac{d\phi}{\sqrt{V_{eff}(\phi_m)-V_{eff}(\phi)}}
\label{eqyphig}
\end{equation}
with as before $\xi=\sqrt{\kappa/(2\mathcal{E}_c)}$ denotes 
the process zone scale.
In the regime $\ell\gg 1$ and $\delta\gg 1$ (or equivalently $\phi_m\ll 1$)
the phase field profile on the process zone scale is independent of
$\delta$
\begin{equation}
\frac{y}{\xi}=\int_{0}^{\phi}\frac{d\phi}{\sqrt{1-g(\phi)}}
\end{equation}
This is not true 
for $\phi$ comparable to $\phi_m$ (in the regime $\ell\gg 1$ and $\delta\gg 1$)
where Eq.~(\ref{eqyphig})
simplifies to
\begin{equation}
\frac{y}{\xi}= \phi_m \int_{1}^{\phi/\phi_m}
\frac{\rho^{\sigma/2} d\rho}{\sqrt{\rho^{\sigma}-1}}
\label{eqyphi}
\end{equation}
In the same regime $\phi\sim \phi_m$, the strain field can be written
\begin{equation}
u=\sqrt{\frac{\kappa}{\lambda+2\mu}}\frac{\phi_m^{1-\sigma/2}}{\sqrt{a}}
\int_{1}^{\phi/\phi_m}
\frac{ d\rho}{\rho^{\sigma/2}\sqrt{\rho^{\sigma}-1}}=\frac{\Delta}{C_{\sigma}}
\int_{1}^{\phi/\phi_m}
\frac{ d\rho}{\rho^{\sigma/2}\sqrt{\rho^{\sigma}-1}}
\label{eqyu}
\end{equation}
Comparison of Eq.~(\ref{eqyphi}) and Eq.~(\ref{eqyu}) shows that the variation
of $u$ is comparable to the total strain
$\Delta$ [Eq.~(\ref{eqd5}]
on a scale
$\phi_m \xi$ much smaller than the process zone length. 
More precisely, for different strains
$\Delta$, the different scaled strain profiles $u/\Delta$ are given
by a unique function of $y/(\phi_m \xi)$.

\section{Variational equations and self-adjointness of linearized operators}
\label{var.app}

The equilibrium phase-field equations considered in this paper are
Euler-Lagrange equations coming from the variation of an energy density
$\mathcal{E}$ where $\mathcal{E}$ depends on a set of fields $\psi_{\alpha}$
(here the elastic displacements and a scalar phase-field) and their spatial
derivatives $\partial_j\psi_{\alpha}$,
\be
\frac{\partial \mathcal{E}}{\partial\psi_{\alpha}} -
\partial_j \frac{\partial \mathcal{E}}{\partial[\partial_j\psi_{\alpha}]}=0
\label{el.eq}
\ee
We show here that quite generally for this type of equations, the allied
linear
operator is self-adjoint, a property that we used for obtaining the explicit
formulae of section \ref{sol.sec}.
 
Linearization of Eq.~(\ref{el.eq}) 
around a solution $\psi_{\alpha}^{(0)}$ produces the linearized operator
$\mathcal{L}$. It is defined  by its action on a set of functions $v_{\beta}$
 as 
\begin{eqnarray}
\mathcal{L}_{\alpha}[\{v_{\beta}\}]=\frac{\partial^2 \mathcal{E}}{\partial\psi_{\alpha}\partial\psi_{\beta}}
v_{\beta} &+& \frac{\partial^2 \mathcal{E}}{\partial\psi_{\alpha}
\partial[\partial_j\psi_{\beta}]} \partial_j v_{\beta} -
\partial_j \left\{\frac{\partial^2 
\mathcal{E}}{\partial\psi_{\beta}\partial[\partial_j\psi_{\alpha}]} v_{\beta}
\right\}
\nonumber\\
&-& \partial_j \left\{
\frac{\partial^2\mathcal{E}}{\partial[\partial_k\psi_{\beta}]\partial[\partial_j
\psi_{\alpha}]} \partial_kv_{\beta}
\right\}
\label{linel.eq}
\end{eqnarray}

Now it is easily seen using integration by parts 
that 
for two arbitrary sets
of differentiable functions $v_{\beta}$ and $w_{\alpha}$, one has
\be
\int \!\!dx\ w_{\alpha} \mathcal{L}_{\alpha}[\{v_{\beta}\}]=
\int \!\! dx\ v_{\alpha} \mathcal{L}_{\alpha}[\{w_{\beta}\}] +\mathrm{boundary\ terms}
\label{sa.eq}
\ee
The relation clearly holds separately for the first and last term on the
r.~h.~s. of Eq.~\ref{linel.eq} and comes from the interchange of the
second and third term on using integration by parts.
Thus, 
the linear operator
$\mathcal{L}_{\alpha}$ is self-adjoint for the usual flat 
measure (here simply $dx\equiv dx_1 dx_2$ on the plane).

\section{Explicit computations of zero-modes and solvability integrals.}
\label{sol.app}
For the convenience of the reader, we provide below some details of
our computations of  solvability
integrals and of the Eshelby tensor line integrals.

For plane strain, the explicit form (\ref{km}) of the stress distribution
near a crack tip is conveniently obtained from the Airy function $\chi$
which satisfies the biharmonic equation
\be
\nabla^2 (\nabla^2 \chi)=0
\label{bhe}
\ee
In polar coordinates $(r,\Theta)$, $\chi$ is related to the strain tensor by
\be
\sigma_{rr}=\frac{1}{r^2}\partial^2_{\Theta\Theta}\chi+
\frac{1}{r}\partial_{r}\chi, \sigma_{\Theta\Theta}=\partial^2_{rr}\chi
\ \mathrm{and}\ \sigma_{r\Theta}=\frac{1}{r^2}\partial_{\Theta}\chi -
\frac{1}{r}\partial^2_{r\Theta}\chi
\label{airypol}
\ee 

For a crack along the $x$-axis, with its tip at $x=0$, the Airy function is 
determined by
Eq.~(\ref{bhe}), together with zero traction boundary conditions on the 
fracture lips, $\sigma_{rr}=\sigma_{r\Theta}=0$
for $\Theta=\pm\pi$. The most singular possibility
compatible with a bounded elastic energy reads, in polar coordinates,
\be
\chi= \frac{r^{3/2}}{3}\left\{ \frac{K_1}{\sqrt{2\pi}}
[ 3 \cos(\frac{\Theta}{2})+
\cos(\frac{3\Theta}{2})]-\frac{K_2}{\sqrt{2\pi}}[\sin(\frac{\Theta}{2})+
\sin(\frac{3\Theta}{2})]
\right\}
\ee 
The dominant divergent forms of the stress distribution follow 
by differentiation with the help of Eq.~(\ref{airypol}),
\begin{eqnarray}
\!\!\!\!\!\sigma_{rr}&=&\frac{K_1}{\sqrt{2\pi r}}[\frac{5}{4}\cos(\frac{\Theta}{2})-
\frac{1}{4}\cos(\frac{3\Theta}{2})]-\frac{K_2}{\sqrt{2\pi r}}
[\frac{5}{4}\sin(\frac{\Theta}{2})-\frac{3}{4}\sin(\frac{3\Theta}{2})]
\nonumber
\\
\!\!\!\!\!\sigma_{\Theta\Theta}&=&\frac{K_1}{\sqrt{2\pi r}}[\frac{3}{4}
\cos(\frac{\Theta}{2})+
\frac{1}{4}\cos(\frac{3\Theta}{2})]-\frac{K_2}{\sqrt{2\pi r}}
[\frac{3}{4}\sin(\frac{\Theta}{2})+\frac{3}{4}\sin(\frac{3\Theta}{2})]
\nonumber
\\
\!\!\!\!\!\sigma_{r\Theta}&=&\frac{K_1}{\sqrt{2\pi r}}[\frac{1}{4}
\sin(\frac{\Theta}{2})+
\frac{1}{4}\sin(\frac{3\Theta}{2})]+\frac{K_2}{\sqrt{2\pi r}}
[\frac{1}{4}\cos(\frac{\Theta}{2})+\frac{3}{4}\cos(\frac{3\Theta}{2})]
\label{str}
\end{eqnarray}
The relation between the strain and stress tensors (Eq.~(\ref{st.def}))
and integration give the allied displacement field,
\be
u_{i}=\frac{1}{4\mu}\sqrt{\frac{r}{2\pi}}\,\left[K_1\, d^I_i +
K_2\, d^{II}_i \right], \ i=r,\Theta
\label{disf}
\ee

The  mode I crack tip displacement functions $ d^I_i$ are given by

\begin{eqnarray}
d^I_{r}&=&
(5-8\nu) \cos(\frac{\Theta}{2})-\cos(\frac{3\Theta}{2})
\nonumber
\\
d^I_{\theta}&=&(-7+8\nu) \sin(\frac{\Theta}{2})+
\sin(\frac{3\Theta}{2})
\label{u1}
\end{eqnarray}
The corresponding  mode II functions are
\begin{eqnarray}
d^{II}_{r}&=&(-5+8\nu) \sin(\frac{\Theta}{2})+3\sin(\frac{3\Theta}{2})
\nonumber
\\
d^{II}_{\Theta}&=&(-7+8\nu) \cos(\frac{\Theta}{2})+3
\cos(\frac{3\Theta}{2})
\label{u2}
\end{eqnarray}

These expressions allow the explicit evaluation of the different
integrals of sections \ref{sol.sec} and \ref{esh.sec}.

{\em i) Solvability integrals.}

For a vectorial field $\mathbf{u}=u_r \mathbf{e}_r +
u_{\Theta} \mathbf{e}_{\Theta}$
the two components of the $x$-translation field
 $\mathbf{u}^{(x)}\equiv\partial_x \mathbf{u}$
are
\begin{eqnarray}
u^{(x)}_r&=&\cos(\Theta)\,\partial_ru_r-\frac{\sin(\Theta)}{r}\,
\partial_{\Theta}u_r+\frac{\sin(\Theta)}{r}\,u_{\Theta}\nonumber\\
u^{(x)}_{\Theta}&=&\cos(\Theta)\,\partial_ru_{\Theta}-\frac{\sin(\Theta)}{r}
\,\partial_{\Theta}u_{\Theta}-\frac{\sin(\Theta)}{r}\,u_{r}
\label{xtf}
\end{eqnarray}
With these formulae, one can compute
the two components of the
x-translation field $\mathbf{u}^{(x;I)}$ associated to the mode I 
the displacement field
[Eq.~(\ref{disf},\ref{u1})],
\begin{eqnarray}
u^{(x;I)}_r&=&\frac{K_1}{8\mu\sqrt{2\pi r}}\,[(7-8\nu)\cos(3\Theta/2)-3\cos(\Theta/2)]
\nonumber\\
u^{(x;I)}_{\Theta}&=&\frac{K_1}{8\mu\sqrt{2\pi r}}\,
[(-5+8\nu)\sin(3\Theta/2)+3\sin(\Theta/2)]
\label{ux1}
\end{eqnarray}
The corresponding strain tensor reads
\begin{eqnarray}
u_{rr}^{(x;I)}=&\partial_r u_r^{(x;I)}&=-\frac{K_1}{16\mu\sqrt{2\pi}\, r^{3/2}}[(7-8\nu)
\cos(3\Theta/2)-3\cos(\Theta/2)]\nonumber\\
u_{\Theta\Theta}^{(x;I)}=&\frac{1}{r}[u_{r}^{(x;I)}+
\partial_{\Theta} u_{\Theta}^{(x;I)}]&=
-\frac{K_1}{16\mu\sqrt{2\pi}\, r^{3/2}}
[(1-8\nu)\cos(3\Theta/2)+3\cos(\Theta/2)]\nonumber\\
u_{r\Theta}^{(x;I)}=&\frac{1}{2}[\partial_{r}u_{\Theta}^{(x;I)}+
\partial_{\Theta} u_{r}^{(x;I)}]&=-
\frac{3 K_1}{16\mu\sqrt{2\pi}\, r^{3/2}}[\sin(3\Theta/2)+\sin(\Theta/2)]
\label{urrx}
\end{eqnarray}
The two needed components of the allied stress tensor follows
\begin{eqnarray}
\sigma_{rr}^{(x;I)}&=&\frac{K_1}{8\sqrt{2\pi}\, r^{3/2}}\, [-7
\cos(3\Theta/2)+3\cos(\Theta/2)]\nonumber\\
\sigma_{r\Theta}^{(x;I)}&=&-\frac{3K_1}{8\sqrt{2\pi}\, r^{3/2}}\,
[\sin(3\Theta/2)+\sin(\Theta/2)]
\label{stx}
\end{eqnarray}
With these explicit expressions, one can evaluate the solvability
integrals for the perturbed displacement field $u_i^{(1)}$,
\begin{equation}
u_{i}^{(1)}=\frac{1}{4\mu}\sqrt{\frac{r}{2\pi}}\,\left[\delta K_1\, d^I_i +
\delta K_2\, d^{II}_i \right], \ i=r,\Theta
\label{disfp}
\end{equation}
With
Eq.~(\ref{str},\ref{ux1}) and (\ref{u1},\ref{stx}),
 one obtains for the two integrals, 
\begin{eqnarray}
\int ds_j u_i^{(1)}\sigma_{ij}^{(x;I)}&=&\int ds_j \left[
\frac{1}{4\mu}\sqrt{\frac{r}{2\pi}} \delta K_1\, d_i^{I}\ 
\sigma_{ij}^{(x;I)}\right]
=\frac{K_1\,\delta K_1}{8\mu}
[5-6\nu]
\nonumber\\
\int ds_j u_i^{(x;I)}\sigma_{ij}^{(1)}&=&
\frac{K_1\delta K_1}{8\mu}
[-3+2\nu]
\label{sint1}
\end{eqnarray}
where $\sigma_{ij}^{(1)}$ is the perturbation of
the stress tensor corresponding to $u_{i}^{(1)}$. It is given by
the mode I part of
Eq.~(\ref{str}) with $K_1$ and $K_2$ replaced by
$\delta K_1$ and $\delta K_2$.
For symmetry reasons, only the
 mode I
part of $u_{i}^{(1)}$ and $\sigma_{ij}^{(1)}$ contribute to the integrals. 
Substracting the two Eqs.~(\ref{sint1}), one finally obtains
\begin{equation}
\int ds_j [u_i^{(x;I)}\sigma_{ij}^{(1)}-u_i^{(1)}\sigma_{ij}^{(x;I)}]=
-\frac{K_1\,\delta K_1}{\mu}[1-\nu]
\label{normx}
\end{equation}
which is equation (\ref{scxsimp}) of the main text.

The corresponding expressions for 
a y-translation field $\mathbf{u}^{(y)}\equiv\partial_y \mathbf{u}$
associated to a vectorial field $\mathbf{u}=u_r \mathbf{e}_r +
u_{\Theta} \mathbf{e}_{\Theta}$
are
\begin{eqnarray}
u^{(y)}_r&=&\sin(\Theta)\,\partial_ru_r+\frac{\cos(\Theta)}{r}\,
\partial_{\Theta}u_r-\frac{\cos(\Theta)}{r}\,u_{\Theta}
\nonumber \\
u^{(y)}_{\Theta}&=&\sin(\Theta)\,\partial_ru_{\Theta}+\frac{\cos(\Theta)}{r}
\,\partial_{\Theta}u_{\Theta}+\frac{\cos(\Theta)}{r}\,u_{r}
\end{eqnarray}
This gives for the two components of
the y-translation field $\mathbf{u}^{(y;I)}$
associated to the mode I displacement field
of Eq.~(\ref{u1})
\begin{eqnarray}
u^{(y;I)}_r&=&\frac{K_1}{8\mu\sqrt{2\pi r}}\,[(7-8\nu)\sin(3\Theta/2)-\sin(\Theta/2)]
\nonumber\\
u^{(y;I)}_{\Theta}&=& \frac{K_1}{8\mu\sqrt{2\pi r}}\,
[(5-8\nu) \cos(3\Theta/2)-\cos(\Theta/2)]
\label{uy1}
\end{eqnarray}
with the corresponding strain tensor
\begin{eqnarray}
u_{rr}^{(y;I)}&=&-\frac{K_1}{16 \mu\sqrt{2\pi} r^{3/2}}\,
[(7-8\nu) \sin(3\Theta/2)-
\sin(\Theta/2)]\nonumber\\
u_{\Theta\Theta}^{(y;I)}&=&- \frac{K_1}{16 \mu\sqrt{2\pi} r^{3/2}}\,
[(1-8\nu)\sin(3\Theta/2)+
\sin(\Theta/2)]\nonumber\\
u_{r\Theta}^{(y;I)}&=&\frac{K_1}{16 \mu\sqrt{2\pi} r^{3/2}}
[3\cos(3\Theta/2)+\cos(\Theta/2)]
\label{urry}
\end{eqnarray}
and stress tensor
\begin{eqnarray}
\sigma_{rr}^{(y;I)}&=&\frac{K_1}{8\sqrt{2\pi}\,r^{3/2}}[-7
\sin(3\Theta/2)+\sin(\Theta/2)]\nonumber\\
\sigma_{r\Theta}^{(y;I)}&=&\frac{K_1}{8\sqrt{2\pi}\,r^{3/2}} [
3\cos(3\Theta/2)+\cos(\Theta/2)]
\end{eqnarray}
This gives for the two integrals of interest
\begin{eqnarray}
\int ds_j u_i^{(y;I)}\sigma_{ij}^{(1)}&=&\frac{K_1 K_2}{8\mu}
[5-6\nu]
\nonumber\\
\int ds_j u_i^{(1)}\sigma_{ij}^{(y;I)}&=&\frac{K_1 K_2}{8\mu}
[-3+2\nu]
\label{sint2}
\end{eqnarray}
where only $\sigma_{ij}^{(1)}$ and $u_i^{(1)}$, the mode II 
parts of the perturbed
the stress tensor $\sigma_{ij}^{(1)}$ and displacements fields
$u_i^{(1)}$ (Eq.~(\ref{disfp}) contribute to the integrals.

One finally obtains for the difference of the two integrals
of Eq.~(\ref{sint2})
\begin{equation}
\int ds_j [u_i^{(y;I)}\sigma_{ij}^{(1)}-u_i^{(1)}\sigma_{ij}^{(y;I)}]=
\frac{K_1\, K_2}{\mu}[1-\nu]
\label{normy}
\end{equation}
which is equation (\ref{scysimp}) of the main text.

{\em ii) Components of the configurational force on the crack tip}

The two line integrals [Eq.~(\ref{fconf})]
giving the two components of the configurational force on the crack tip
can be directly evaluated with the help of the above results.
\be
f_i^{(conf)}=\int_{A\rightarrow B} \!\!\! ds \,T^E_{ij}\,n_j=
\int_{A\rightarrow B} \!\!\! ds\ [\mathcal{E}_{strain} n_i -
\sigma_{jk} u_k^{(i)}]
\ee
where $\mathbf{u}^{(i)}\equiv \partial_i\mathbf{u}$ is the translation field in the direction $i\, (i=x,y)$,  
with mode I and mode II components included
\be
u_k^{(i)}=u_k^{(i,I)}+u_k^{(i,II)}
\ee
The mode I components $u_k^{(i,I)}$ are given by Eq.~(\ref{ux1}) and 
(\ref{uy1}). A similar computation gives their mode II components
with the help of Eqs.~(\ref{disf},\ref{u2}) and (\ref{xtf})
\begin{eqnarray}
u^{(x;II)}_r&=&\frac{K_2}{8\mu\sqrt{2\pi r}}\,[(-7+8\nu)\sin(3\Theta/2)+\sin(\Theta/2)]
\nonumber\\
u^{(x;II)}_{\Theta}&=& \frac{K_2}{8\mu\sqrt{2\pi r}}\,
[(-5+8\nu) \cos(3\Theta/2)+\cos(\Theta/2)]
\label{utx2}
\end{eqnarray}
and
\begin{eqnarray}
u_r^{(y,II)}&=&\frac{K_2}{8\mu\sqrt{2\pi r}}\,[(7-8\nu)\cos(3\Theta/2)
+5\cos(\Theta/2)]
\nonumber\\
u_{\Theta}^{(y,II)}&=&\frac{K_2}{8\mu\sqrt{2\pi r}}\,[(-5+8\nu)\sin(3\Theta/2)
-5\sin(\Theta/2)]
\end{eqnarray}
With these formulae and the stress tensor expression Eq.~(\ref{str}), one
obtains
\begin{eqnarray}
\int_{-\pi}^{+\pi}\!\! r d\Theta\ [\sigma_{rr} u_r^{(x)}+
\sigma_{r\Theta} u_{\Theta}^{(x)}]&=&\frac{1}{8\mu}[K_1^2 (-3+2\nu)+K_2^2 (-5+6\nu)]
\label{f1s}
\nonumber\\
\int_{-\pi}^{+\pi}\!\! r d\Theta\ [\sigma_{rr} u_r^{(y)}+
\sigma_{r\Theta} u_{\Theta}^{(y)}]&=&\frac{1}{4\mu} K_1 K_2 (3-2\nu)
\label{f2s}
\end{eqnarray}

Furthermore, with $\mathcal{E}_{strain}=(\sigma_{ij}\sigma_{ij}-\nu
\sigma_{ii}\sigma_{jj})$ and the stress tensor expression Eq.~(\ref{str}),
one obtains
\begin{eqnarray}
\int_{-\pi}^{+\pi}\!\! r d\Theta\ \cos(\Theta) \mathcal{E}_{strain}&=&
\frac{1}{8\mu}(K_1^2-K_2^2)(1-2\nu)
\label{f1e}
\\
\int_{-\pi}^{+\pi}\!\! r d\Theta\ \sin(\Theta) \mathcal{E}_{strain}&=&
-\frac{1}{4\mu} K_1 K_2 (1-2\nu)
\label{f2e}
\end{eqnarray}
 
Substraction of Eq.~(\ref{f1s}) from Eq.~(\ref{f1e}) gives
the usual expression of $F_1^{(conf)}$,
\be
F_1^{(conf)}=\int_{-\pi}^{+\pi}\!\! d\Theta\ [\cos(\Theta) \mathcal{E}_{strain}-
\sigma_{rr}u^{(x)}_r
-\sigma_{r\Theta} u^{(x)}_{\Theta}]=\frac{1-\nu}{2\mu}(K_1^2+K_2^2)
\label{f21eapp}
\ee
as given in Eq.~(\ref{f21e}) in the main text.

The other component of the configurational force $F_2^{(conf)}$
is similarly obtained by substracting Eq.~(\ref{f2s}) from Eq.~(\ref{f2e}),
\be
F_2^{(conf)}=\int_{-\pi}^{+\pi}\!\! d\Theta\ [\sin(\Theta) \mathcal{E}_{strain}-
\sigma_{rr} u^{(y)}_r
-\sigma_{r\Theta} u^{(y)}_{\Theta}]=-\frac{1-\nu}{\mu} K_1 K_2
\label{f22eapp}
\ee
which is Eq.~(\ref{f22e}) of the main text.

\end{document}